\documentclass[
               reprint, twocolumn,
               preprintnumbers,
               amsmath,
               amssymb,
               showpacs,
               aip,jcp,
               superscriptaddress]{revtex4-2}

 \usepackage{ucs}
 \usepackage[utf8x]{inputenc}
 \usepackage{latexsym}
 \usepackage{amsfonts}
 \usepackage{amsmath}
 \usepackage{graphicx}
 \usepackage{color}
 \usepackage{bm}
\graphicspath{ {./figs/} }
\begin{document}

\author{Sudip Sasmal}
\email[e-mail: ]{sudip.sasmal@pci.uni-heidelberg.de}
\affiliation{Theoretische Chemie,
             Physikalisch-Chemisches Institut,
             Universität Heidelberg,
             Im Neuneheimer Feld 229, 69120 Heidelberg, Germany}

\author{Markus Schröder}
\email[e-mail: ]{markus.schroeder@pci.uni-heidelberg.de}
\affiliation{Theoretische Chemie,
             Physikalisch-Chemisches Institut,
             Universität Heidelberg,
             Im Neuneheimer Feld 229, 69120 Heidelberg, Germany}

\author{Oriol Vendrell}
\email[e-mail: ]{oriol.vendrell@uni-heidelberg.de}
\affiliation{Theoretische Chemie,
             Physikalisch-Chemisches Institut,
             Universität Heidelberg,
             Im Neuneheimer Feld 229, 69120 Heidelberg, Germany}
\affiliation{Interdisciplinary Center for Scientific Computing,
             Universität Heidelberg,
             Im Neuneheimer Feld 205, 69120 Heidelberg, Germany}
%\author{coauthor 1}
%\email[e-mail: ]{coauthor@couathor.com}
%\affiliation{somewhere}

\begin{abstract}
We propose an approach to represent the second-quantized electronic Hamiltonian in a compact
sum-of-products (SOP) form. The approach is based on the canonical polyadic decomposition (CPD) of
the original Hamiltonian projected onto the sub-Fock spaces formed by groups of spin orbitals.
The algorithm for obtaining the canonical polyadic form starts from
an exact sum-of-products, which is then optimally compactified using an alternating
least-squares procedure.
%This procedure aims at mitigating the quartic scaling of the number of terms in the
%Hamiltonian with respect to the number of spin orbitals.
%The primary objective of obtaining a compact sum-of-products form is to facilitate the
%treatment of
%application of this method to
%
We discuss the relation of this specific SOP with related forms, namely the Tucker
format and the matrix product operator often used in conjunction with matrix product
states.
We benchmark the method on the electronic dynamics of an excited water molecule,
trans-polyenes, and the charge migration in glycine upon
inner-valence ionization. The quantum dynamics are performed 
with the multilayer multi-configuration time-dependent Hartree method
in second quantization representation (MCTDH-SQR). Other
methods based on tree-tensor Ans{\"a}tze may profit from
this general approach.
\end{abstract}

\title{Compact sum-of-products form of the molecular electronic Hamiltonian based on
       canonical polyadic decomposition}

\date{\today}

% --- Begin article ---

\maketitle

%%%%%%%%%%%%%%%%%%%%%%%%%%%%%%%%%%%%%%%%%%%%%%%%%%%%%%%%%%%%%%%%%%%%%%%%
%%%%%%%%%%%%%%%%%%%%%%%%%%%%%%%%%%%%%%%%%%%%%%%%%%%%%%%%%%%%%%%%%%%%%%%%
\section{Introduction} \label{sec:introduction}
%%%%%%%%%%%%%%%%%%%%%%%%%%%%%%%%%%%%%%%%%%%%%%%%%%%%%%%%%%%%%%%%%%%%%%%%
The multiconfiguration time-dependent Hartree (MCTDH)
~\cite{mey90:73,man92:3199,bec00:1,mey03:251,mey09:book}
and its multilayer generalization (ML-MCTDH)
~\cite{wan03:1289,man08:164116,Ven11:44135,mey12:351,wan15:7951}
are highly efficient methods for simulating the quantum dynamics of
nuclear degrees of freedom in high-dimensional systems. In its original
form, the MCTDH \emph{Ansatz} cannot describe systems of indistinguishable
particles, as it assumes a product form of the underlying single particle
functions (SPFs), thereby failing to account for the proper symmetry of
these indistinguishable particles. However, it is possible to construct the
multiconfiguration wavefunction using Slater determinants and permanents as
the basis to address fermionic and bosonic systems, respectively.
These specialized theories are referred to as MCTDH for fermions (MCTDH-F)
~\cite{cai05:12712,alo07:154103,hoc11:084106,sat13:023402,lod20:011001}
and bosons (MCTDH-B)~\cite{alo07:154103,alo08:33613}.
A unified version of the two theories has been established, employing
a non-symmetric core tensor to connect mixtures of different types of
indistinguishable particles~\cite{kroe13:63018}.
However, a limitation of such descriptions arises from the combinatorial
increase in the number of configurations within the electronic/bosonic
subsystem as the number of particles and single-particle functions grows.
Additionally, the requirement of antisymmetry or symmetry hinders their
further decomposition into smaller-rank tensors. Consequently, when
dealing with the same type of indistinguishable particles, both the
MCTDH-B and MCTDH-F approaches are incompatible with the multilayer
extension of the MCTDH framework.
\par
A fundamentally different approach for describing systems of indistinguishable
particles is the use of the second quantization representation. Wang and Thoss
introduced and applied this approach within the context of MCTDH, naming
it MCTDH in SQR (MCTDH-SQR)~\cite{wan09:024114}.
In this representation, the state of the system is described using the
occupation number representation, which corresponds to the occupation of
specific sets of spin-orbitals. The symmetry of the indistinguishable particles
is expressed through the creation and annihilation operators that operate on
the system's state. Since the
occupation of individual spin-orbitals is treated as analogous
to a coordinate in the traditional MCTDH formulation, the degrees of freedom (DOFs) become
distinguishable, making it straightforward to construct a multilayer
Ansatz for the wavefunction.
\par
The MCTDH-SQR method has found applications in various domains after its introduction,
including solving the impurity problem in non-equilibrium dynamical mean-field
theory~\cite{Bal15:45136} and addressing quantum transport in molecular
junctions~\cite{Wan11:244506,wan13:134704,wan13:7431}
and quantum dots~\cite{Wil13:045137,Wil14:205129}.
Additionally, Manthe and Weike developed an MCTDH-SQR approach based on
time-dependent optimal orbitals, known as the
MCTDH-oSQR method~\cite{man17:064117,wei20:034101}.
These applications primarily dealt with model Hamiltonians
~\cite{wan09:024114,Bal15:45136,man17:064117,wei20:034101,Wan11:244506,
wan13:134704,wan13:7431,Wil13:045137,Wil14:205129}.
More recently, we extended the MCTDH-SQR method to describe non-adiabatic
dynamics in molecular systems. This extension is based on the second-quantized
representation of the electrons and the first-quantized representation of the
nuclear coordinates, while also providing expressions for the non-adiabatic
coupling matrix elements within this combined representation~\cite{Sas20:154110}.
In this formalism, the non-adiabatic effects are considered within
the time-evolving electronic subsystem coupled with the dynamics of the nuclei,
thus bypassing the explicit determination of potential energy surfaces and
non-adiabatic couplings and offering an alternative
to the conventional group Born-Oppenheimer (BO) approximation.

Nonetheless, a significant challenge remains when applying the
MCTDH-SQR method to
\emph{ab initio} studies of large molecular systems, namely the
unfavourable scaling of the number of terms in the
electronic SQR Hamiltonian, which increases %exponentially
with
the fourth power of the number of spin-orbitals.
The density matrix renormalization group
(DMRG)~\cite{Whi92:2863,Whi93:10345,sch05:259}
formalism, which shares a tensor representation framework similar
to ML-MCTDH, mitigates this scaling by representing
the Hamiltonian as a matrix product operator
(MPO)~\cite{Sch11:96,Cha11:465,McC07:P10014,Cha16:014102,Kel15:244118,Yan14:283},
compatible with the matrix product state (MPS)~\cite{Sch11:96, McC07:P10014}
structure of the wavefunction.
In contrast, a comparably concise representation of the electronic Hamiltonian
compatible with the sum-of-products (SOP)--based MCTDH algorithm is not yet available.
%Within the SOP form, a high-dimensional operator is represented as a sum of products of low-dimensional operators acting on the primitive (or physical) degrees of freedom.
%In other words, these low-dimensional operators must entirely exist within
%the space spanned by the corresponding physical degree of freedom.
In the SOP form, a high-dimensional operator is represented as a sum of products of operators. Every
operator in each product acts
within the space of the corresponding primitive (or physical) degree
of freedom.

When treating high-dimensional nuclear dynamics problems,
obtaining such a SOP representation of the operator is a major challenge:
while the kinetic energy operator is often in this form, this is not true for the potential energy surface. A wide range
of methods have therefore been developed to transform general potential energy surfaces
into SOP format. These methods encompass approaches such as
POTFIT~\cite{bec00:1,jaec96:7974,jaec98:3772},
multigrid POTFIT~\cite{Pel13:014108},
Monte Carlo POTFIT~\cite{Sch17:064105}, as well as the
multilayer extension of POTFIT~\cite{ott14:14106,Ott18:116},
and Monte Carlo canonical polyadic decomposition (MCCPD)~\cite{Sch20:024108}.
Additionally, some algorithms utilize neural network techniques~\cite{Man05:5295,Man06:084109,Man06:194105,Koc14:021101,
She15:144701,Pra16:174305,Pra16:158}.
Note also  that there are alternative MCTDH implementations
that do not require the SOP form of the operator - for example, the correlation discrete variable representation
of Manthe~\cite{Har05:064106}
and the collocation-MCTDH method of
Wodraszka and Carrington~\cite{Wod18:044115}.

Recently, we introduced an approach based on the Tucker decomposition, hence similar
to POTFIT, to represent
a compact SOP form of the second quantized electronic Hamiltonian,
where the term {\it compact} refers to a smaller number of SOP terms compared
to the original SQR Hamiltonian~\cite{Sas22:134102}.
Unfortunately, this approach suffers from
a poor scaling due to the presence of
a core tensor acting on
a direct product of operator subspaces.
In the present work, we introduce and benchmark an approach based on canonical
polyadic decomposition (CPD),
which serves two purposes:
first, it mitigates the quartic scaling of the electronic Hamiltonian
with respect to the number of spin-orbitals.
Second, 
compared to the Tucker format,
CPD does not feature a core tensor of expansion coefficients
on top of an orthogonal basis of one-particle operators. Instead,
the products of the expansion are independent of each other
offering, potentially, a much more favourable scaling. 
As we discuss below, the MPO form often used in DMRG lies
conceptually in between two limiting cases: the
Tucker format and CPD.

The paper is organized as follows.
Section~\ref{sec:theory:sop} discusses the general strategy to write the
SQR electronic Hamiltonian in SOP form.
Section~\ref{sec:theory:dof} reviews the choice of DOF in the MCTDH-SQR
formalism.
Section~\ref{sec:theory:compact} introduces strategies that
lead to compact SOP forms of the electronic SQR Hamiltonian, in particular
CPD, and its connection to the MPO form.
Secs~\ref{sec:res:h2o} and ~\ref{sec:res:polyene} and ~\ref{sec:res:glycine}
present and discuss numerical results on H$_2$O, trans-C$_8$H$_{10}$, and
glycine, respectively.
The scaling of various coefficients related to the ML-MCTDH-SQR method
is briefly discussed in Section~\ref{sec:scaling}.
Finally, a summary and conclusions are provided in Section~\ref{sec:conclusions}.
%%%%%%%%%%%%%%%%%%%%%%%%%%%%%%%%%%%%%%%%%%%%%%%%%%%%%%%%%%%%%%%%%%%%%%%%

%%%%%%%%%%%%%%%%%%%%%%%%%%%%%%%%%%%%%%%%%%%%%%%%%%%%%%%%%%%%%%%%%%%%%%%%
\section{Theory}
%%%%%%%%%%%%%%%%%%%%%%%%%%%%%%%%%%%%%%%%%%%%%%%%%%%%%%%%%%%%%%%%%%%%%%%%
\subsection{Sum-of-products form of the electronic Hamiltonian}  \label{sec:theory:sop}
%%%%%%%%%%%%%%%%%%%%%%%%%%%%%%%%%%%%%%%%%%%%%%%%%%%%%%%%%%%%%%%%%%%%%%%%
The \emph{ab initio} electronic Hamiltonian in the second quantization framework reads
%%%%%%%%%%%%%%%%%%%%%%%%%%%%%%%%%%%%%
\begin{align}
 \hat{H} = \sum_{ij} h_{ij} \hat{a}_i^{\dagger}\hat{a}_j + \frac{1}{2}
    \sum_{ijkl}
    v_{ijkl}\hat{a}_i^{\dagger}\hat{a}_j^{\dagger}\hat{a}_l\hat{a}_k ,
    \label{eq:ham_spinorb}
\end{align}
where
\begin{align}
  h_{ij} = \langle \phi_i(1)| -\frac{1}{2}\nabla_1^2 - \sum_{A=1}^M \frac{Z_A}{r_{_{1A}}} | \phi_j(1) \rangle , \\
  v_{ijkl} = \langle \phi_i(1) \phi_j(2) | \frac{1}{r_{12}} | \phi_k(1) \phi_l(2) \rangle ,
\end{align}
%%%%%%%%%%%%%%%%%%%%%%%%%%%%%%%%%%%%%
are the one- and two-body integrals, involving the
spin orbitals $\phi_i$, respectively.
The $\hat{a}_i$ and $\hat{a}_i^{\dagger}$ correspond to the annihilation
and creation operators that annihilate and create an electron in the $i$-th
spin orbitals, respectively, satisfy the fermionic commutation relations
%%%%%%%%%%%%%%%%%%%%%%%%%%%%%%%%%%%%%
\begin{align}
 \label{eq:anticom}
 \{\hat{a}_i, \hat{a}_j^{\dagger} \} =
    \hat{a}_i\hat{a}_j^{\dagger} + \hat{a}_j^{\dagger}\hat{a}_i 
    = \delta_{ij} , \\
 \label{eq:antic4}
  \{\hat{a}_i^{\dagger}, \hat{a}_j^{\dagger} \} = \{\hat{a}_i, \hat{a}_j \} = 0.
\end{align}
%%%%%%%%%%%%%%%%%%%%%%%%%%%%%%%%%%%%%
Although the electronic Hamiltonian
in Eq.~\ref{eq:ham_spinorb} seems like already being in
a SOP form, the anti-commutation relations of the fermionic creation and 
annihilation operators in Eqs.~\ref{eq:anticom} and \ref{eq:antic4}  lead to
the accumulation of a phase factor $S_s = \sum_{k=1}^{s-1} n_k$ depending
on the occupation of all spin-orbitals before the $s$-th position
for $\hat{a}_s^\dagger$ acting on a Fock-space configuration
(the same is true for $\hat{a}_s$)~\cite{Fetter2003}
%%%%%%%%%%%%%%%%%%%%%%%%%%%%%%%%%%%%%
\begin{align}
    \label{eq:opfermi}
    \hat{a}_s^\dagger |n_1, n_2, \dots , n_M \rangle  =
            \hat{a}_s^\dagger (\hat{a}_1^\dagger)^{n_1}
                      (\hat{a}_2^\dagger)^{n_2} \cdots
                      (\hat{a}_M^\dagger)^{n_M}
                      |\mathrm{0}\rangle \nonumber \\
         =
            (-1)^{S_s} (\hat{a}_1^\dagger)^{n_1}
                       (\hat{a}_2^\dagger)^{n_2}    \cdots
                       [\hat{a}_s^\dagger(\hat{a}_s^\dagger)^{n_s}] \cdots
                      (\hat{a}_M^\dagger)^{n_M}
                       |\mathrm{0}\rangle.
\end{align}
%%%%%%%%%%%%%%%%%%%%%%%%%%%%%%%%%%%%%
This phase factor makes the creation and annihilation operator
nonlocal, i.e., the operator $\hat{a}_s^{(\dagger)}$ acts beyond
its index $s$ and therefore the electronic SQR Hamiltonian is,
in general, not in the SOP form with respect
to the primitive degrees of freedom.
Wang and Thoss resolved this issue by mapping the fermionic operators onto
equivalent spin operators~\cite{wan09:024114}.
Formally, this mapping consists of applying the inverse Jordan-Wigner (JW)
transformation to the fermionic field operators and effectively transforming the
fermionic Hamiltonian into an equivalent spin-chain Hamiltonian~\cite{jor28:631}.
The equivalent spin-$1/2$ chain Hamiltonian after the JW transformation reads~\cite{Sas20:154110}
%%%%%%%%%%%%%%%%%%%%%%%%%%%%%%%%%%%%%
\begin{align}
  \label{eq:ham_el_jw}
    \hat{H} = &
    \sum_{ij} h_{_{ij}}
        \left(
          \prod_{q=a+1}^{b-1} \hat{\sigma}^z_q
        \right) \hat{\sigma}_i^{+} \hat{\sigma}_j^{-} \nonumber\\
     + &  \frac{1}{2}\sum_{ijkl} v_{_{ijkl}}
          \left( \prod_{q=a+1}^{b-1} \hat{\sigma}^z_q
              \prod_{q^\prime=c+1}^{d-1} \hat{\sigma}^z_{q^\prime}
          \right) \nonumber\\
       &   \mathrm{sgn}(j-i)\mathrm{sgn}(l-k)
    \hat{\sigma}_i^{+} \hat{\sigma}_j^{+} \hat{\sigma}_l^{-} \hat{\sigma}_k^{-}.
\end{align}
%%%%%%%%%%%%%%%%%%%%%%%%%%%%%%%%%%%%%
where
$\hat{\sigma}_i^{+} = \frac{1}{2}\left( \hat{\sigma}_i^{x}+i\hat{\sigma}_i^{y} \right)$,
$\hat{\sigma}_i^{-} = \frac{1}{2}\left( \hat{\sigma}_i^{x}-i\hat{\sigma}_i^{y} \right)$, and
$\hat{\sigma}_k^{z}$ are the standard spin ladder operators with Pauli matrices
$\bm{\sigma}^{x}$, $\bm{\sigma}^{y}$ and $\bm{\sigma}^{z}$.
Here the indices $(a,b,c,d)$ correspond to the $(i,j,k,l)$ indices, but ordered from
smaller to larger and the $\mathrm{sgn}$ function is defined as
%%%%%%%%%%%%%%%%%%%%%%%%%%%%%%%%%%%%%
\begin{align}
    \mathrm{sgn}(x) =
        \begin{cases}
            1, & \text{if } x\geq 0\\
           -1, & \text{otherwise}.
        \end{cases}
\end{align}
%%%%%%%%%%%%%%%%%%%%%%%%%%%%%%%%%%%%%
The operators $\hat{\sigma}_\kappa^{+}$, $\hat{\sigma}_\kappa^{-}$
and $\hat{\sigma}^z_\kappa$ act now locally on $\kappa$-th spin-$\frac{1}{2}$
basis function and their matrix representation reads
\begin{align}
 \label{eq:mat_s_dof}
 \bm{\sigma}^{+} =
 \begin{pmatrix}
    0 & 0 \\
    1 & 0 \\
 \end{pmatrix} ;\,\,\,\,
 \bm{\sigma^{-}} =
 \begin{pmatrix}
    0 & 1 \\
    0 & 0 \\
 \end{pmatrix} ;\,\,\,\,
 \bm{\sigma^{z}} =
\begin{pmatrix}
    1 & 0 \\
    0 & -1 \\
\end{pmatrix} .
\end{align}
Clearly, the electronic Hamiltonian written in Eq.~\ref{eq:ham_el_jw} is in the
SOP form with respect to the primitive DOFs (spin-$\frac{1}{2}$ basis).
%%%%%%%%%%%%%%%%%%%%%%%%%%%%%%%%%%%%%
\subsection{Spin and Fock space degree of freedom} \label{sec:theory:dof}
%%%%%%%%%%%%%%%%%%%%%%%%%%%%%%%%%%%%%
%%  Figure: MCTDH WF tree
\begin{figure}[t]
\begin{center}
\includegraphics[height=6cm]{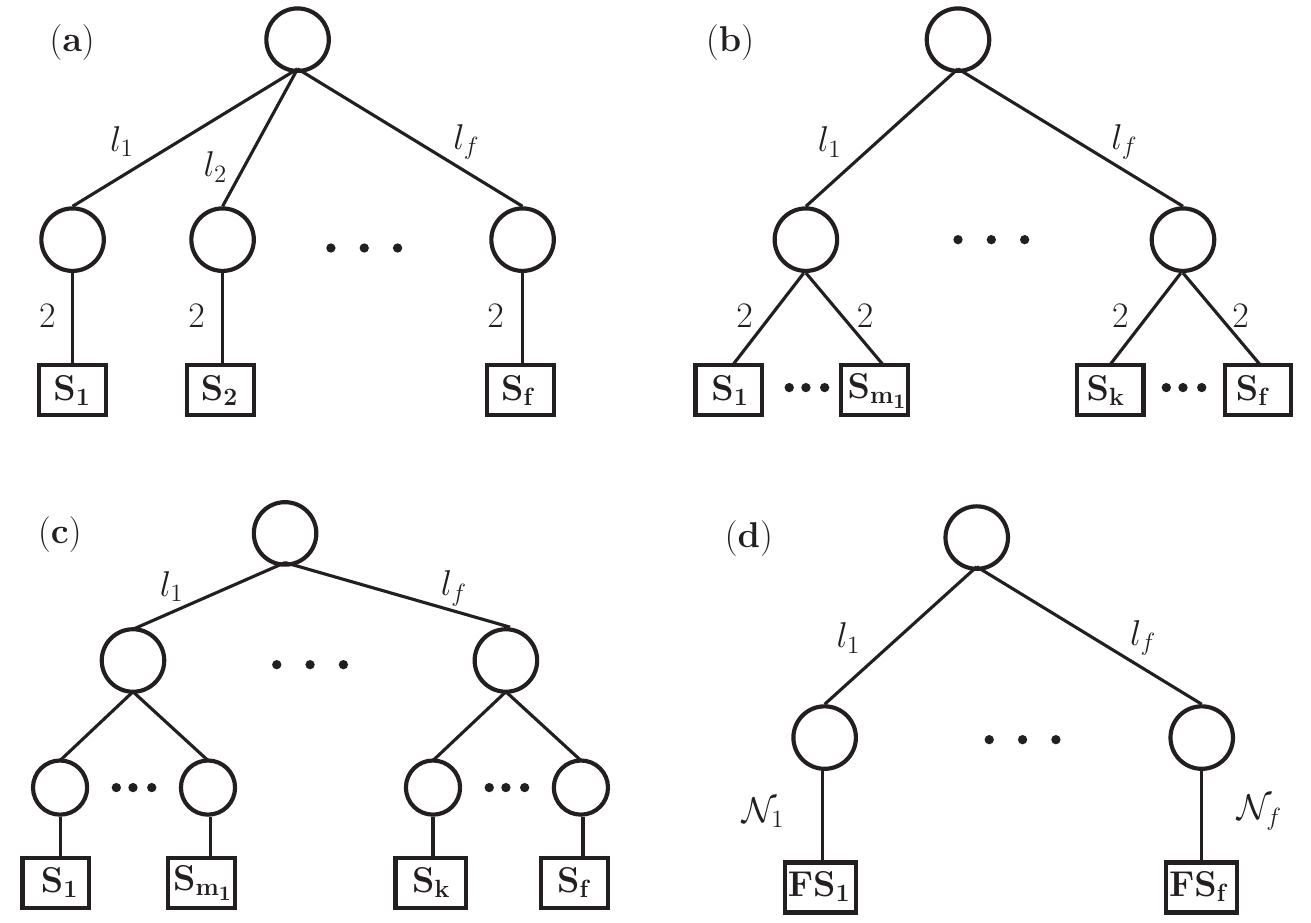}
\caption{Tree structures for the various MCTDH-SQR wavefunction \emph{Ansätze}:
         (a) Normal MCTDH wavefunction tree, in which the spin degree of
             freedoms (S-DOF) are considered as primitive DOFs.
         (b) MCTDH wavefunction tree with mode combination, in which
             S-DOFs are considered as primitive DOFs.
         (c) ML-MCTDH wavefunction tree, in which
             S-DOFs are considered as primitive DOFs.
         (d) MCTDH wavefunction tree, in which the Fock space of a group
             of spin orbitals is considered as single DOF (FS-DOF).
         $l_\kappa$ denotes number of SPF of the $\kappa$-th mode and
         $\mathcal{N}_{\kappa}$ is the number of configuration of
         $\kappa$-th FS-DOF.
    }
\label{fig:mctdhwfn}
\end{center}
\end{figure}
%%%%%%%%%%%%%%%%%%%%%%%%%%%%%%%%%%%%%
In the above formulation
there are two limiting MCTDH-SQR wavefunction \emph{Ansätze}:
Each spin degree of freedom (S-DOF) is
described by either (i) one or (ii) two time-dependent SPFs.
The former corresponds to a
time-dependent Hartree (TDH) wavefunction with a single
Hartree product and the latter corresponds to the exact wavefunction consisting of
2$^{M}$ configurations, where M is the number of S-DOF.
Thus, the limiting case (i) leads to a poor description of a correlated state
and the latter leads to an exact formulation, which becomes quickly unaffordable as the number
of S-DOF (i.e., spin orbitals) increases.
Therefore, the only practical
way to apply the MCTDH-SQR method is to create groups of S-DOFs, either
through mode combination or through its multi-layer generalization (ML-MCTDH).
Figures~\ref{fig:mctdhwfn} (a), (b), and (c) show the normal MCTDH wavefunction,
MCTDH wavefunction with mode combination, and ML-MCTDH wavefunction, respectively,
where S-DOFs are used as primitive DOFs.
\par
For $M$ spin orbitals there are 2$^M$ Fock states.
On the other hand, one can divide the total Fock space into $f$ sub-Fock
spaces by grouping the M spin orbitals into $f$ groups
($m_1$, $m_2$, $\cdots$, $m_f$)
%%%%%%%%%%%%%%%%%%%%%%%%%%%%%%%%%%%%%
\begin{align}
 \mathcal{F}(M) = \mathcal{F}_1(m_1) \otimes \mathcal{F}_2(m_2)
 \otimes \dots \otimes \mathcal{F}_f(m_f)
 \label{eq:fsdof_fock}
\end{align}
%%%%%%%%%%%%%%%%%%%%%%%%%%%%%%%%%%%%%
where $\mathcal{F}_\kappa(m_\kappa)$ denotes the sub-Fock space of 
the $\kappa$-th FS-DOF (consists of $m_\kappa$ S-DOFs) with
$\mathcal{N}_\kappa$ (=2$^{m_\kappa}$) Fock states and
%%%%%%%%%%%%%%%%%%%%%%%%%%%%%%%%%%%%%
\begin{align}
 M = \sum_{\kappa=1}^{f} m_\kappa.
\end{align}
%%%%%%%%%%%%%%%%%%%%%%%%%%%%%%%%%%%%%
Now, one can represent the configurations of the sub-Fock space
$\mathcal{F}_\kappa(m_\kappa)$ as a new primitive DOF. We refer to this representation
of the primitive DOF as Fock space DOF (FS-DOF).
Fig.~\ref{fig:mctdhwfn} (d) shows the MCTDH wavefunction involving FS-DOF
as primitive DOF.
In the FS-DOF formalism, one needs to transform the primitive operator string described in
the Eq.~\ref{eq:ham_el_jw} into new primitive matrix operators acting onto the sub-Fock
spaces of their corresponding FS-DOF.

In the FS-DOF representation, the state of the system is described by \emph{kets} $|i_1,\ldots,i_f\rangle$, where
$i_\kappa$ corresponds to the $i$-th configuration of the $\kappa$-th FS-DOF. One can think of the
$i_\kappa$ configuration indices as indexing Euclidean basis vectors within each degree of freedom.
The configurations
within a FS-DOF can correspond, in general, to different electron occupation numbers.
A matrix element of the total Hamiltonian is then represented as
\begin{align}
    \label{eq:ham_fsdof}
    H_{z_1,\ldots,z_f} = H_{i_1,\ldots,i_f}^{j_1,\ldots,j_f} =
    \langle j_1,\ldots,j_f| \hat{H} | i_1,\ldots,i_f \rangle,
\end{align}
where $z_\kappa = (i_\kappa, j_\kappa)$ are double indices indicating
the \emph{bra}- and \emph{ket}-side configurations within each FS-DOF.
This electronic Hamiltonian is very sparse and, clearly, it can only be explicitly
constructed and diagonalized for small systems. In the following, thus, we will discuss
the construction of sum-of-products forms of the Hamiltonian, Eq.~\ref{eq:ham_fsdof},
and their use within the framework of MCTDH-SQR.
%%%%%%%%%%%%%%%%%%%%%%%%%%%%%%%%%%%%%%%%%%%%%%%%%%%%%%%%%%%%%%%%%%%%%%%%

%%%%%%%%%%%%%%%%%%%%%%%%%%%%%%%%%%%%%%%%%%%%%%%%%%%%%%%%%%%%%%%%%%%%%%%%
\subsection{Compact sum-of-products form of the electronic Hamiltonian}
     \label{sec:theory:compact}
%%%%%%%%%%%%%%%%%%%%%%%%%%%%%%%%%%%%%%%%%%%%%%%%%%%%%%%%%%%%%%%%%%%%%%%%

Although the Hamiltonian, Eq. \ref{eq:ham_el_jw}, has already the desired sum-of-products
form to be used in ML-MCTDH-SQR calculations, it contains too many terms for being
practically applicable in all but the most simple molecular systems.
For this reason, much effort in our group has been devoted to devising
procedures to arrive at more compact forms of the \emph{ab initio} SQR Hamiltonian.
% COMMENT REVIEWER 1
%\textcolor{blue}{
%where the term {\it compact} refers to a smaller number of SOP terms.
%}
%
In this section, we introduce various SOP forms
of the SQR electronic Hamiltonian and discuss their general properties and how
they are related.
The numerically efficient construction of a compact SQR form of the operator
is discussed below in Section~\ref{sec:theory:sop2sop}.

%%%%%%%%%%%%%%%%%%%%%%%%%%%%%%%%%%%%%%%%%%%%%%%%%%%%%%%%%%%%%%%%%%%%%%%%
\subsubsection{Tucker decomposition (T-SQR)} \label{sec:theory:tsqr}
%%%%%%%%%%%%%%%%%%%%%%%%%%%%%%%%%%%%%%%%%%%%%%%%%%%%%%%%%%%%%%%%%%%%%%%%
The Tucker SQR (T-SQR) form approximates the electronic Hamiltonian tensor
($H_{z_1,\ldots,z_f}$) in the FS-DOF primitive basis as~\cite{Sas22:134102}
%%%%%%%%%%%%%%%%%%%%%%%%%%%%%%%%%%%%%
\begin{align}
 H_{z_1,\ldots,z_f}^{\text{T}} \approx \sum_{l_1}^{n_1} \cdots \sum_{l_f}^{n_f}
                g_{l_1,\cdots,l_f}
                \prod_{\kappa=1}^{f} [\mathbf{O}_{l_\kappa}^{(\kappa)}]_{z_\kappa},
 \label{eq:elham_tucker}
\end{align}
%%%%%%%%%%%%%%%%%%%%%%%%%%%%%%%%%%%%%
where $g_{l_1,\cdots,l_f}$ are the elements of
the Tucker core tensor with rank $(n_1, n_2, \cdots, n_f)$ and
$\mathbf{O}_{l_{\kappa}}^{(\kappa)}$ is the $l_{\kappa}$-th single particle operator (SPO)
acting of the $\kappa$-th FS-DOF.
The operator matrices $\mathbf{O}_{l_{\kappa}}^{(\kappa)}$ are defined in the
sub-Fock space ($\mathcal{F}_\kappa(m_\kappa)$) formed by the $\kappa$-th
FS-DOF.
When expanded as vectors, these elements form an orthonormal set,
%%%%%%%%%%%%%%%%%%%%%%%%%%%%%%%%%%%%%
\begin{align}
    \label{eq:SPO_Tucker}
    \sum_{z_\kappa} [\mathbf{O}_{l_\kappa}^{(\kappa)}]_{z_\kappa} \cdot
    [\mathbf{O}_{p_\kappa}^{(\kappa)}]_{z_\kappa} = \delta_{l_\kappa, p_\kappa}.
\end{align}
%%%%%%%%%%%%%%%%%%%%%%%%%%%%%%%%%%%%%
One can obtain the core tensor and SPO matrices by minimizing the
sum-of-squares error function
%%%%%%%%%%%%%%%%%%%%%%%%%%%%%%%%%%%%%%
 \begin{align}
 \label{eq:tuckerr}
     \mathcal{L}^2 = 
     \sum_{z_1} \cdots \sum_{z_f}
     \Bigg( &
           H_{z_1,\ldots,z_f} - H_{z_1,\ldots,z_f}^{(T)}
     \Bigg) ^2,
 \end{align}
%%%%%%%%%%%%%%%%%%%%%%%%%%%%%%%%%%%%%%
where $H_{z_1,\ldots,z_f}^{\text{T}}$ takes the form of Eq.~(\ref{eq:elham_tucker}).
In Ref.~\citenum{Sas22:134102}, the T-SQR form was obtained using
\textit{higher-order orthogonal iteration} (HOOI)~\cite{kol09:455} procedure
in the \texttt{TensorLy}~\cite{tensorly} python library.

Alternatively,
as done in the context of the \texttt{POTFIT} algorithm,\cite{bec00:1,jaec96:7974,jaec98:3772}
 the SPOs can also be obtained by diagonalizing the single particle reduced density matrix of the form
%%%%%%%%%%%%%%%%%%%%%%%%%%%%%%%%%%%%%%
 \begin{align}
  \label{eq:rdm}
  \rho_{z_\kappa,z_\kappa^{\prime}}^{(\kappa)} =&
  \sum_{z_1} \cdots \sum_{z_{\kappa-1}} \sum_{z_{\kappa+1}} \cdots \sum_{z_f} \nonumber\\
  & H_{z_1,\ldots,z_{\kappa-1},z_\kappa,z_{\kappa+1},\ldots,z_f} H_{z_1,\ldots,z_{\kappa-1},z_\kappa^{\prime},z_{\kappa+1},\ldots,z_f}.
 \end{align}
%%%%%%%%%%%%%%%%%%%%%%%%%%%%%%%%%%%%%%
Defining composite indices
%%%%%%%%%%%%%%%%%%%%%%%%%%%%%%%%%%%%%%
\begin{align}
 \label{eq:compindx}
 Z^\kappa &= (z_1, \ldots, z_{\kappa-1}, z_{\kappa+1}, \ldots z_f),\nonumber\\
 Z &= (z_1, \ldots, z_f) = z_\kappa Z^\kappa   ,
\end{align}
%%%%%%%%%%%%%%%%%%%%%%%%%%%%%%%%%%%%%%
the reduced density matrix reads
%%%%%%%%%%%%%%%%%%%%%%%%%%%%%%%%%%%%%%
\begin{align}
 \rho_{z_\kappa,z_\kappa^{\prime}}^{(\kappa)} =&
 \sum_{Z^\kappa} H_{z_\kappa Z^\kappa} H_{z_\kappa^{\prime} Z^\kappa} \nonumber\\
 =& \sum_{I^\kappa} \sum_{J^\kappa}
 H_{i_\kappa I^\kappa}^{j_\kappa J^\kappa} H_{i_\kappa^{\prime} I^\kappa}^{j_\kappa^{\prime} J^\kappa} .
 \label{eq:density:tucker}
\end{align}
%%%%%%%%%%%%%%%%%%%%%%%%%%%%%%%%%%%%%%
The eigenvectors of the reduced density matrix fulfill Eq.~\ref{eq:SPO_Tucker},
and thus can be used as SPOs, whereas the eigenvalues give an estimate of
how important the corresponding SPOs are in the expansion given in
Eq.~(\ref{eq:elham_tucker}).

Note that, for the two-dimensional case,
the eigenvectors of the density matrix are the optimal SPOs in the sense
of an optimal Schmidt decomposition,
and the core tensor can be obtained by
overlapping the original tensor with the SPOs~\cite{bec00:1,jaec96:7974,jaec98:3772}.
For higher dimensions, one can still use the eigenvectors as SPOs
(as these are very close to the optimal ones)
and one only needs to iterate a few times to obtain the core tensor.

Although the T-SQR form introduced in Ref.~\citenum{Sas22:134102} has proven too
costly to fit to become a practicable alternative, it is quite enlightening
to examine the structure of the reduced density matrix in Eq.~(\ref{eq:density:tucker})
and the corresponding
eigenvectors as obtained from the SQR electronic Hamiltonian.

As seen by inspecting the second line of Eq.~(\ref{eq:density:tucker}),
configurations $|i_\kappa, I^\kappa \rangle$ and $|j_\kappa, J^\kappa \rangle$
must belong to the same Hilbert space $\mathcal{H}(N)$ with particle number $N$, while
configurations $|i_\kappa^{\prime}, I^\kappa \rangle$ and $|j_\kappa^{\prime}, J^\kappa \rangle$
must both belong to the same Hilbert space $\mathcal{H}(N^\prime)$ with $N^\prime$ particles,
%%%%%%%%%%%%%%%%%%%%%%%%%%%%%%%%%%%%%%
\begin{align}
 \{|i_\kappa, I^\kappa \rangle, |j_\kappa, J^\kappa \rangle\} &\in \mathcal{H}(N) \nonumber \\
 \{|i_\kappa^{\prime}, I^\kappa \rangle, |j_\kappa^{\prime}, J^\kappa \rangle\} &\in \mathcal{H}(N^{\prime}).
 \end{align}
%%%%%%%%%%%%%%%%%%%%%%%%%%%%%%%%%%%%%%
It follows that 
% where $\mathcal{H}(N)$ and $\mathcal{H}(N^{\prime})$ are the Hilbert space of particle number $N$ and
%$N^{\prime}$. This follows
 \begin{align}
  N_{i_\kappa} + N_{I^\kappa} & = N_{j_\kappa} + N_{J^\kappa} = N \\
  N_{i_\kappa^\prime} + N_{I^\kappa} & = N_{j_\kappa^\prime} + N_{J^\kappa}  = N^\prime
\end{align}
and therefore
 \begin{align}
  N_{i_\kappa} - N_{j_\kappa} = N_{i_\kappa^\prime} - N_{j_\kappa^\prime} \\
  \label{eq:z-kappa}
  \Delta N_{z_\kappa} = \Delta N_{z_\kappa^\prime}
\end{align}
%%%%%%%%%%%%%%%%%%%%%%%%%%%%%%%%%%%%%%
where, $\Delta N_{z_\kappa}$ is the particle number difference between configurations $|j_\kappa\rangle$ and 
$|i_\kappa \rangle$
associated to the combined index $z_\kappa = (i_\kappa, j_\kappa)$
in the $\kappa$-th FS-DOF.
%The above identity is the manifestation of the fact that Hilbert spaces consist of configurations with different
%electron number lies in orthogonal space (i.e., Hamiltonian commutes with particle number operator).

Due to (\ref{eq:z-kappa}), the density matrix in Eq.~(\ref{eq:density:tucker})
factorizes in blocks of different $\Delta N_{z_\kappa}$.
Moreover, since the electronic Hamiltonian cannot connect overall configurations that differ by
the occupation of more than two spin-orbitals,
 \begin{align}
  \Delta N_{z_\kappa} =\Delta N_{z_\kappa^{\prime}} \le 2.
 \end{align}
This results in the block diagonal structure of the density matrix
consisting at most of 13 blocks corresponding to
\begin{align}
\label{eq:blocks}
\Delta N_{z_\kappa} = \{(0,0), (\pm 1, 0), (0, \pm 1), (\pm 2, 0), (0, \pm 2),
(\pm 1, \pm1) \}
\end{align}
which are labeled as $\{ S_0, A_1^{\pm}, B_1^{\pm} A_2^{\pm}, B_2^{\pm}, C_2^{\pm,\pm}\}$, respectively.
The parenthesis's first and second numbers in (\ref{eq:blocks})
indicate the change in alpha and beta electrons, respectively.
%%%%%%%%%%%%%%%%%%%%%%%%%%%%%%%%%%%%%%
%%  Figure: density matrix
\begin{figure}[t]
\begin{center}
\includegraphics[height=8cm]{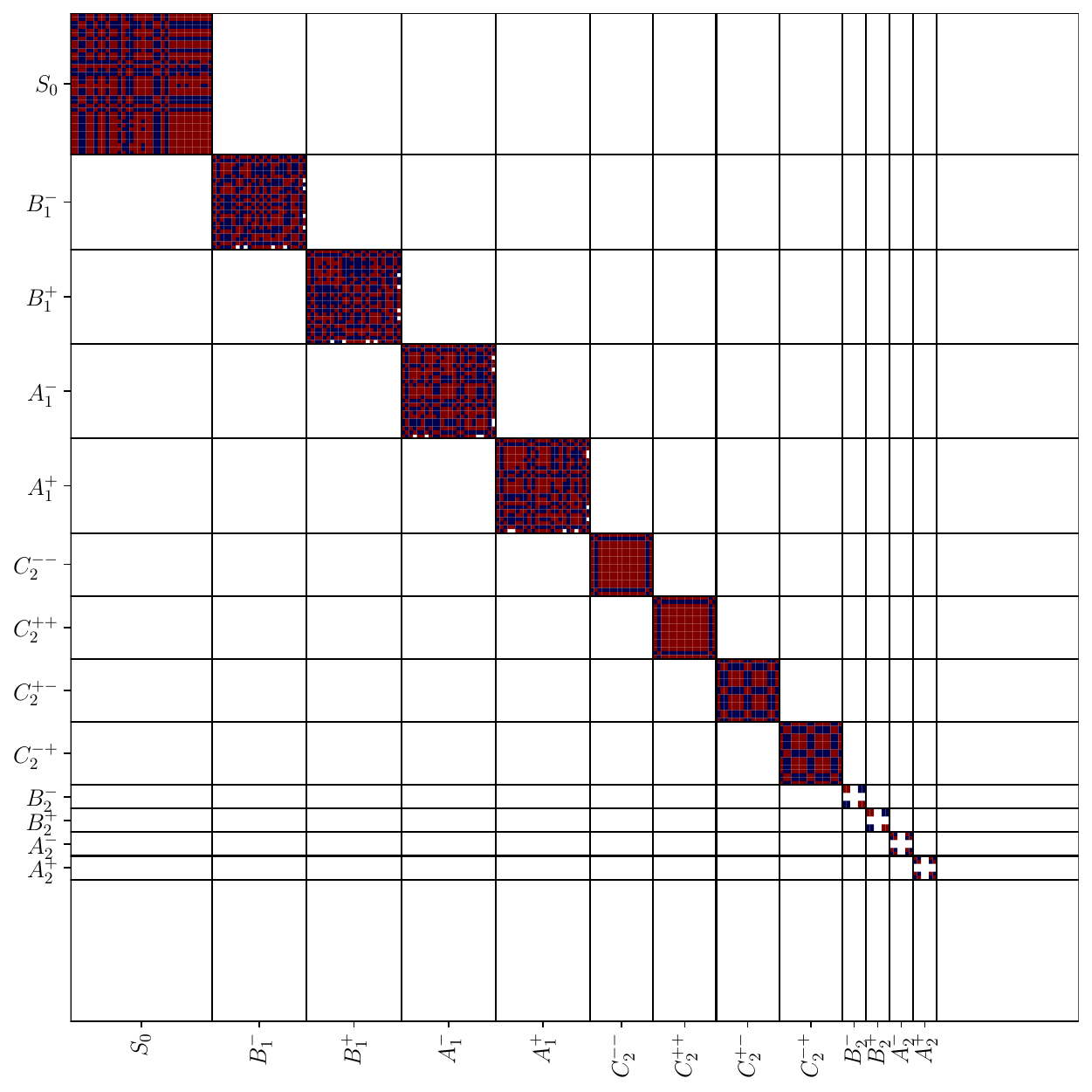}
\caption{The block diagonal structure of the reduced density matrix with 13 blocks.
    }
\label{fig:density}
\end{center}
\end{figure}
%%%%%%%%%%%%%%%%%%%%%%%%%%%%%%%%%%%%%
%%%%%%%%%%%%%%%%%%%%%%%%%%%%%%%%%%%%%%
%%  Figure: density matrix
\begin{figure}[t]
\begin{center}
\includegraphics[height=6cm]{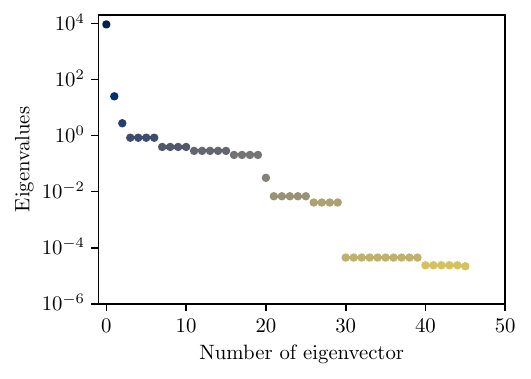}
\caption{The eigenvalues of the density matrix.
    }
\label{fig:den_eval}
\end{center}
\end{figure}
%%%%%%%%%%%%%%%%%%%%%%%%%%%%%%%%%%%%%
%%%%%%%%%%%%%%%%%%%%%%%%%%%%%%%%%%%%%%
%%  Figure: density matrix eigenvalues
\begin{figure}[t]
\begin{center}
\includegraphics[height=7cm]{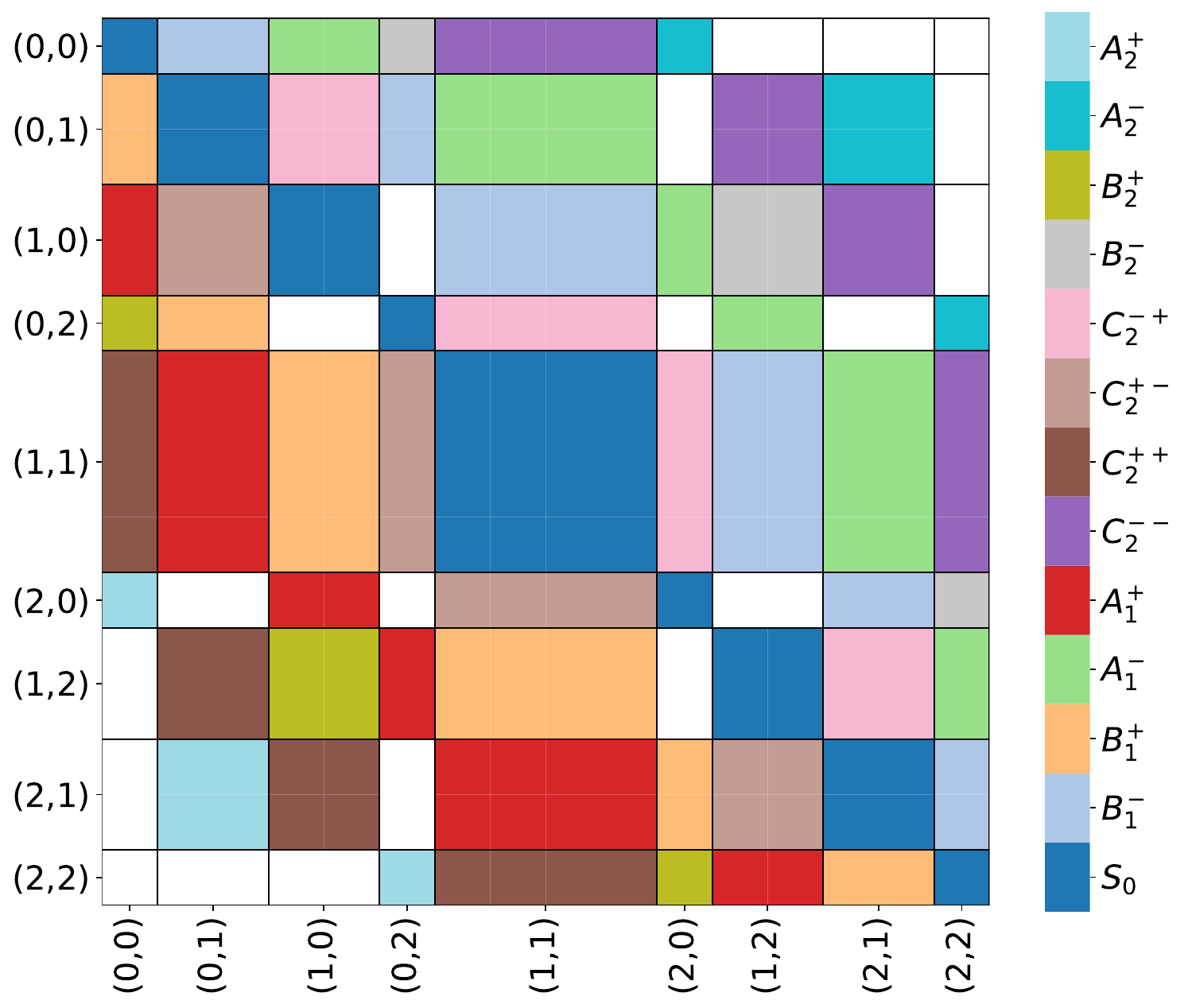}
\caption{The sparse structure of the SPOs obtained by diagonalizing density matrix.
         Here, we show the SPOs of an FS-DOF consists of 4 spin (2 alpha and 2 beta)
         orbitals with 16 configurations. All 13 basic types of SPOs are shown in different
         colors. The non-zero elements of different SPO matrices are only within the
         blocks of the corresponding color.
    }
\label{fig:tsqr_spo}
\end{center}
\end{figure}
%%%%%%%%%%%%%%%%%%%%%%%%%%%%%%%%%%%%%

%To illustrate the block diagonal structure of the reduced density matrix, we considered LiH molecule with STO-3G basis.
%The four occupied spin MOs are grouped into one FS-DOF and four unoccupied $A_1$ spin MOs are grouped into another FS-DOF.
Figure~\ref{fig:density} shows, as an example, the block diagonal structure of the reduced density matrix for the FS-DOF consisting of
four occupied spin MOs of LiH using the STO-3G basis.
The eigenvalues of the density matrix, shown in Fig.~\ref{fig:den_eval}, provide insight into the significance of the corresponding SPO and offer an estimate of how many should be utilized.
In this numerical example, given the rapid decrease in eigenvalues, it is anticipated that numerical convergence can be achieved with fewer (10-15) SPOs.
\par
Due to its block structure, it is not necessary to compute the full density matrix,
potentially a very expensive task,
and indeed in practice one
only needs to evaluate these 13 blocks and diagonalize them separately
to obtain the SPOs.
%As evident from the density matrix, the SPO
%matrices will also have certain sparsity depending on the eigenvector of which block it corresponds.
Additionally, due to the block structure of the operator density matrix,
its eigenvectors are very sparse, and so are the corresponding SPO matrices
obtained by folding the latter into 2-index objects.
The sparse structure of the SPOs is shown in Fig.~\ref{fig:tsqr_spo},
where all 13 types of SPO appear in different colors. The corresponding
operators have non-zero elements only within the blocks
indicated by the corresponding color.
Each of the unique 13 types of basis SPO
has different physical meaning;\\
(i) $S_0$ SPOs move electrons within different orbitals of the FS-DOF, \\
(ii) $A_1^{+/-}$ SPOs add/remove one alpha electron, respectively, \\
(iii) $B_1^{+/-}$ SPOs add/remove one beta electron, respectively \\
(iv) $A_2^{+/-}$ SPOs add/remove two alpha electrons, respectively \\
(v) $B_2^{+/-}$ SPOs add/remove two beta electrons, respectively \\
(vi) $C_2^{++/--}$ SPOs add/remove one alpha and one beta electron, respectively \\
(vii) $C_2^{+-/-+}$ SPOs add/remove one alpha and remove/add one beta electron. \\

This structure reveals that many core-tensor
elements in a T-SQR operator are zero;
only those terms that conserve the particle number are non-zero.
For example,
in a system with two FS-DOFs, $A_1^{+}$ SPOs acting on
the first FS-DOF can only be paired with
$A_1^{-}$ SPOs acting on the second FS-DOF,
and the core-tensor element for all 12 other pairs
must be zero to conserve the total
particle number.
Hence, the SPOs obtained from diagonalization of the operator density matrix
behave quite similarly to particle creation/annihilation operators
(or in general to ladder operators of some kind) acting on the Euclidian
space defined by FS-DOF configurations.
%%%%%%%%%%%%%%%%%%%%%%%%%%%%%%%%%%%%%%%%%%%%%%%%%%%%%%%%%%%%%%%%%%%%%%%%

%%%%%%%%%%%%%%%%%%%%%%%%%%%%%%%%%%%%%%%%%%%%%%%%%%%%%%%%%%%%%%%%%%%%%%%%
\subsubsection{Canonical polyadic decomposition (CPD-SQR)}  \label{sec:theory:cpsqr}
%%%%%%%%%%%%%%%%%%%%%%%%%%%%%%%%%%%%%%%%%%%%%%%%%%%%%%%%%%%%%%%%%%%%%%%%
%%%%%%%%%%%%%%%%%%%%%%%%%%%%%%%%%%%%%%
%%  Figure: hcpd opmat
\begin{figure}[t]
\begin{center}
\includegraphics[height=7cm]{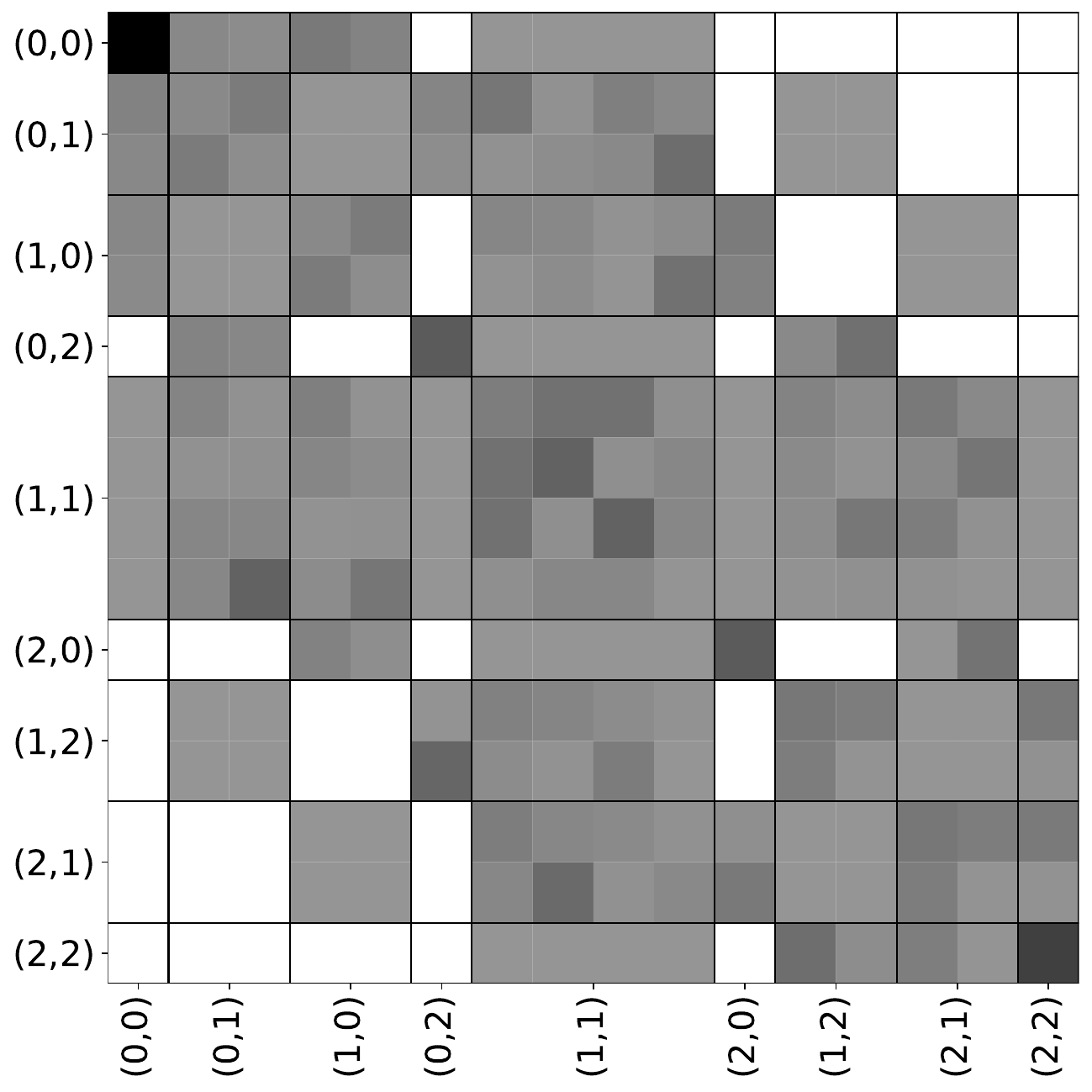}
\caption{Example of a CPD-SQR single particle operator of an FS-DOF consisting of 4 spin
         (2 alpha and 2 beta) orbitals with 16 configurations.
         The single particle operator of a CPD-SQR Hamiltonian does not have
         any particular sparse structure and can have any form demanded by the minimization
         algorithm.
         %Here we show only one SPO of CPD-SQR.
         % Although it looks like the SPO has some sparsity, the other SPOs, in general, do not have such.
    }
\label{fig:cpsqr_spo}
\end{center}
\end{figure}
%%%%%%%%%%%%%%%%%%%%%%%%%%%%%%%%%%%%%
The Tucker form grows exponentially with the number of FS-DOFs
due to the core-tensor~\cite{Sas22:134102}.
%and it is not the most compact sum-of-products representation.
An alternative is the well-known
canonical polyadic decomposition (CPD) also known as CANDECOMP or PARAFAC form
in the literature~\cite{Hit27:164, har70:1, Car70:238, Kie00:105, Sch20:024108}.
The CPD form of the SQR (CPD-SQR) electronic Hamiltonian reads
%%%%%%%%%%%%%%%%%%%%%%%%%%%%%%%%%%%%%
\begin{align}
    \label{eq:candecomp}
    H_{z_1,\ldots,z_f}^{\text{CPD}} = \sum_{r=1}^R c_r \prod_{\kappa=1}^f [\mathbf{X}_r^{(\kappa)}]_{z_\kappa},
\end{align}
%%%%%%%%%%%%%%%%%%%%%%%%%%%%%%%%%%%%%
where $R$ is the rank of the CPD, $c_r$ and $\mathbf{X}_r^{(\kappa)}$ are the
coefficient and SPO of $\kappa$-th FS-DOF in the $r$-th term of the expansion.
The main difference from the Tucker form is that the SPOs acting on the
$\kappa$-th degree of freedom are no longer orthogonal to each
other (cf. Eq.~(\ref{eq:SPO_Tucker})).
%, i.e.,  
%%%%%%%%%%%%%%%%%%%%%%%%%%%%%%%%%%%%%
%\begin{align}
%    \label{eq:SPO_CPD}
%    \sum_{z_\kappa} [\mathbf{X}_l^{(\kappa)}]_{z_\kappa} \cdot %[\mathbf{X}_p^{(\kappa)}]_{z_\kappa}
%    \begin{cases}
%     = 1, & \text{if } l=p \\
%     \ne 0, & \text{otherwise},
%    \end{cases}
%\end{align}
%%%%%%%%%%%%%%%%%%%%%%%%%%%%%%%%%%%%%
%and hence, there is no core tensor.
This additional freedom makes the CPD form a very compact
SOP representation of the tensor.
On the other hand, this also makes the CPD much harder to obtain. 
The numerical challenge is to find both the coefficient and the SPOs, and often the
so-called alternating least squares (ALS) method is used. This is our approach as
outlined in Section~\ref{sec:theory:sop2sop}.

When comparing the structure of the SPOs obtained from T-SQR and CPD-SQR
it becomes quite clear that CPD-SQR can achieve a more compact form of the
sum-of-products operator.
As discussed when considering the density matrix in Fig.~\ref{fig:density},
each of the 13 basic types of SPOs performs a specific task
(e.g. annihilating an $\alpha$-electron from the corresponding FS-DOF),
which is reflected
in the sparse structure of the corresponding matrix representations of
the SPOs, Fig.~\ref{fig:tsqr_spo}.
On the other hand, the SPOs in a compactified CPD-SQR are no longer constrained
by the orthogonality condition and, in general, can assume any form dictated by the
minimization procedures. In particular, they can no longer be ascribed to any of the
previously discussed operator categories, as seen by inspecting Fig.~\ref{fig:cpsqr_spo}.
This additional freedom of the SPOs, not being members of an orthonormal basis
of operators, will prove crucial in achieving a very compact representation
of the SQR operator.
%%%%%%%%%%%%%%%%%%%%%%%%%%%%%%%%%%%%%%%%%%%%%%%%%%%%%%%%%%%%%%%%%%%%%%%%

%%%%%%%%%%%%%%%%%%%%%%%%%%%%%%%%%%%%%%%%%%%%%%%%%%%%%%%%%%%%%%%%%%%%%%%%
\subsubsection{Relation to matrix product operators}
        \label{sec:theory:mpo}
%%%%%%%%%%%%%%%%%%%%%%%%%%%%%%%%%%%%%%%%%%%%%%%%%%%%%%%%%%%%%%%%%%%%%%%%
%%%%%%%%%%%%%%%%%%%%%%%%%%%%%%%%%%%%%%
%%  Figure: sop op
\begin{figure*}[t]
\begin{center}
\includegraphics[height=7cm]{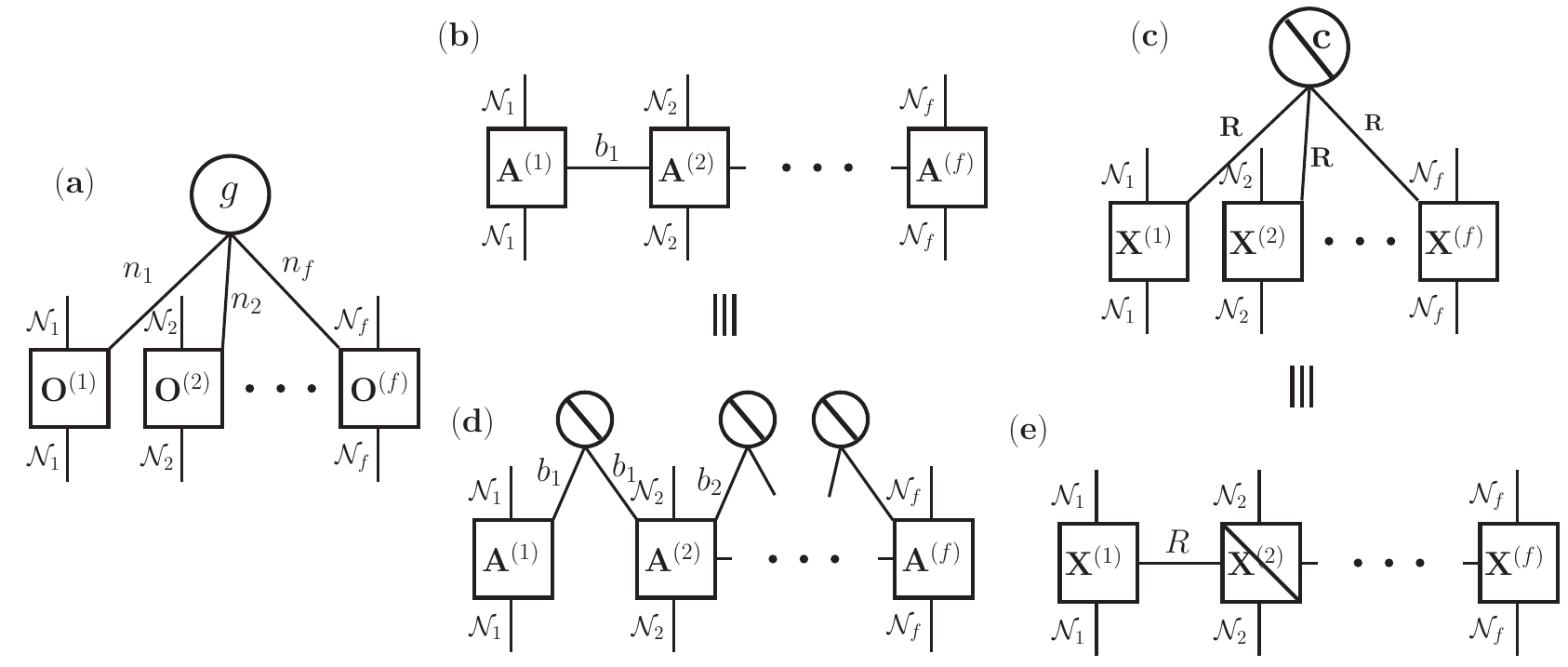}
\caption{Diagrams for various sum-of-products operators.
         The square box and circle are single particle operators and core tensors, respectively.
         The connected lines are contracted indices and disconnected lines are the
         primitive indices. The diagonal line within a box or circle indicates
         the contracted indices connecting SPOs or core tensors are diagonal.
         T-SQR, MPO, and CPD-SQR forms are shown in (a), (b), and (c), respectively.
         Alternative forms of MPO and CPD-SQR are shown in (d) and (e), respectively.
    }
\label{fig:spoop}
\end{center}
\end{figure*}
%%%%%%%%%%%%%%%%%%%%%%%%%%%%%%%%%%%%%
There is yet another SOP form of the operator, the matrix product
operator (MPO)~\cite{Sch11:96,Cha11:465,McC07:P10014,Cha16:014102,Kel15:244118,Yan14:283}.
The MPO form of the electronic Hamiltonian reads
%%%%%%%%%%%%%%%%%%%%%%%%%%%%%%%%%%%%%
\begin{align}
 \label{eq:mpo}
 H_{z_1,\ldots,z_f}^{\text{MPO}} = \sum_{\alpha} [\mathbf{A}_{\alpha_1}^{(1)}]_{z_1} \cdots  [\mathbf{A}_{\alpha_{\kappa-1}, \alpha_{\kappa}}^{(\kappa)}]_{z_\kappa} \cdots [\mathbf{A}_{\alpha_{f-1}}^{(f)}]_{z_f} ,
\end{align}
%%%%%%%%%%%%%%%%%%%%%%%%%%%%%%%%%%%%%
where, $\alpha$ is a combined index of ($\alpha_1, \ldots, \alpha_{f-1}$) and
$\mathbf{A}_{\alpha_{\kappa-1}, \alpha_{\kappa}}^{(\kappa)}$
are operators acting on site $\kappa$, henceforth equivalent to the SPO introduced above
and referred to as such in the following. The non-terminal SPOs 
have two indices, one shared with the tensor to its left and the other with the
tensor to its right side.
%
%Terminal SPOs have just one index, similarly : the leftmost SPO shares the index with
%its right SPO, and the rightmost SPO shares the index with its left SPO.

From the point of view of tensor decomposition, the MPO can be viewed as an intermediate
between the Tucker decomposition and
the CPD. 
The Tucker, MPO, and CPD diagrammatic forms are shown in Fig.~\ref{fig:spoop} (a), (b),
and (c), respectively.
The square boxes and circles indicate SPOs and core tensors, respectively.
The connecting lines correspond to contracted (summed) indices and disconnected lines indicate
primitive indices, i.e., indices of the original tensor.
Diagonal lines within the boxes or circles indicate that the corresponding tensor is
diagonal in the indices connecting SPOs or core tensors (and not the primitive indices).
In Tucker format (Fig.~\ref{fig:spoop}(a)),
the core tensor coefficients $g$ are contraction coefficients multiplying the
SPOs of all FS-DOF, thus the exponential scaling of this form.
In an MPO, only neighboring SPOs are connected (Fig.~\ref{fig:spoop}(b)).
From the perspective of a Tucker form, this can
be viewed as if a two-dimensional diagonal core tensor
(i.e., a diagonal matrix) connects them, as depicted in Fig.~\ref{fig:spoop}(d).
Similarly, from the perspective of a Tucker form the CPD form
(Fig.~\ref{fig:spoop}(c)) can be regarded as possessing a diagonal core tensor, so
all connecting lines count with the same index.
Alternatively, one can view the CPD as an MPO where the $\mathbf{A}_{\alpha_{\kappa-1}, \alpha_{\kappa}}^{(\kappa)}$ tensors are diagonal in the $\alpha$ indices,
as indicated by the diagonal line in the non-terminal diagonal boxes in Fig.~\ref{fig:spoop}(d).

The MPO form of the Hamiltonian may turn out to be very efficient at capturing couplings
between neighboring sites and, as in the matrix product state representation
of the wavefunction, it may be advantageous if the physical
system has a connectivity that is well-suited to the topology of the MPO
tensor representation. On the other hand, a CPD operator is, in principle, free from such restraint. However, the
\emph{ansatz} of the wavefunction determines to a large extent what form
of the Hamiltonian constitutes an
optimal choice from a numerical perspective,
and the MPO form of the Hamiltonian is compatible with the matrix
product states (MPS)
representation of the wavefunction~\cite{Sch11:96,Cha11:465,McC07:P10014,Cha16:014102,Kel15:244118,Yan14:283}.
Our ML-MCTDH implementation works efficiently with general SOPs and, in
principle, it is possible to unfold an MPO and use it with MCTDH.
It is yet to be investigated if the
(ML-)MCTDH \emph{ansatz} and algorithm can profit from the MPO structure
in some way that does not involve its trivial unfolding into the explicit
enumeration of all product terms contained in Eq.~(\ref{eq:mpo}).
This question is beyond the scope of this work.
%On the other hand, the MPO, in its native form, cannot efficiently be used in
%(ML-)MCTDH algorithm as the SPOs share a common index with its neighbor and thus, not in
%a SOP form. However, one can unwind the shared index to make it in a SOP form.
%%%%%%%%%%%%%%%%%%%%%%%%%%%%%%%%%%%%%%%%%%%%%%%%%%%%%%%%%%%%%%%%%%%%%%%%

%%%%%%%%%%%%%%%%%%%%%%%%%%%%%%%%%%%%%%%%%%%%%%%%%%%%%%%%%%%%%%%%%%%%%%%%
\subsection{Obtaining compact CPD-SQR operators via ALS} \label{sec:theory:sop2sop}

%%%%%%%%%%%%%%%%%%%%%%%%%%%%%%%%%%%%%%%%%%%%%%%%%%%%%%%%%%%%%%%%%%%%%%%%

\subsubsection{General approach to finding an optimal CPD}

Before discussing the ALS method in detail, let us introduce a number of quantities.
Following the notation written in Eq.~(\ref{eq:compindx}), let
%%%%%%%%%%%%%%%%%%%%%%%%%%%%%%%%%%%%%
\begin{align}
 [\Omega_r]_Z = \prod_\kappa [\mathbf{X}_r^{(\kappa)}]_{z_\kappa}
\end{align}
%%%%%%%%%%%%%%%%%%%%%%%%%%%%%%%%%%%%%
be the product of all SPOs for one expansion term r and
matrix element $Z=(z_1,z_2,\ldots)$,
then the CPD form
of the Hamiltonian (\ref{eq:candecomp}) can be rewritten as
%%%%%%%%%%%%%%%%%%%%%%%%%%%%%%%%%%%%%
\begin{align}
 H_{Z}^{\text{CPD}} = \sum_r c_r [\Omega_r]_Z .
\end{align}
%%%%%%%%%%%%%%%%%%%%%%%%%%%%%%%%%%%%%
Let us also define the single-hole product of all SPOs
%%%%%%%%%%%%%%%%%%%%%%%%%%%%%%%%%%%%%
\begin{align}
 [\Omega_r^{\kappa}]_{Z^{\kappa}} = \prod_{\kappa^{\prime} \ne \kappa} [\mathbf{X}_r^{(\kappa^{\prime})}]_{z_{\kappa^{\prime}}} .
\end{align}
%%%%%%%%%%%%%%%%%%%%%%%%%%%%%%%%%%%%%
Let us further define the positive semidefinite single-hole overlap matrices
%%%%%%%%%%%%%%%%%%%%%%%%%%%%%%%%%%%%%
\begin{align}
 S_{r,r^{\prime}}^{(\kappa)} = \sum_{Z^\kappa} [\Omega_r^{\kappa}]_{Z^{\kappa}} [\Omega_{r^{\prime}}^{\kappa}]_{Z^{\kappa}} ,
\end{align}
%%%%%%%%%%%%%%%%%%%%%%%%%%%%%%%%%%%%%
which are separable, i.e.,
%%%%%%%%%%%%%%%%%%%%%%%%%%%%%%%%%%%%%
\begin{align}
 S_{r,r^{\prime}}^{(\kappa)} = \prod_{\kappa^{\prime} \ne \kappa} \zeta_{r,r^{\prime}}^{(\kappa^{\prime})},
\end{align}
where
\begin{align}
 \zeta_{r,r^{\prime}}^{(\kappa^{\prime})} = \sum_{z_{\kappa^{\prime}}} [\mathbf{X}_r^{(\kappa^{\prime})}]_{z_{\kappa^{\prime}}} [\mathbf{X}_{r^{\prime}}^{(\kappa^{\prime})}]_{z_{\kappa^{\prime}}},
 \label{eq:zeta}
\end{align}
%%%%%%%%%%%%%%%%%%%%%%%%%%%%%%%%%%%%%
which can be exploited for numerical calculation.
To obtain coefficients and SPOs, one defines a functional $J$ that is to be optimized as
%%%%%%%%%%%%%%%%%%%%%%%%%%%%%%%%%%%%%
\begin{align}
 \mathcal{J} = \sum_{Z} (H_Z - H_Z^{\text{CPD}})^2 + \epsilon \sum_r c_r^2 \sum_Z [\Omega_r]_Z^2 .
\end{align}
%%%%%%%%%%%%%%%%%%%%%%%%%%%%%%%%%%%%%
Here the first term measures the error of the CPD-SQR expansion, Eq.~(\ref{eq:candecomp}), with
respect to the original Hamiltonian, and the second term serves as a regularization term with
regularization parameter $\epsilon$. It is introduced to avoid the appearance of linear dependent terms in the CPD expansion during the optimization. The regularization parameter $\epsilon$ is set to a small number, usually the square root of machine precision, i.e. $\epsilon = 10^{-8}$.
The functional derivative of $\mathcal{J}$ with respect to the SPO of one FS-DOF times
the coefficient yields
%%%%%%%%%%%%%%%%%%%%%%%%%%%%%%%%%%%%%
\begin{align}
 \frac{\delta \mathcal{J}}{\delta c_r  [\mathbf{X}_{r}^{(\kappa)}]_{z_\kappa} } =
 &- 2 \sum_{Z^\kappa} H_Z [\Omega_r^{\kappa}]_{Z^{\kappa}}
 +2 \sum_{r^{\prime}} c_{r^{\prime}}  [\mathbf{X}_{r}^{(\kappa)}]_{z_\kappa} S_{r,r^{\prime}}^{(\kappa)} \nonumber\\
 &+ 2 \epsilon c_r  [\mathbf{X}_{r}^{(\kappa)}]_{z_\kappa} S_{r,r}^{(\kappa)} \longmapsto 0.
 \label{eq:jcp}
\end{align}
%%%%%%%%%%%%%%%%%%%%%%%%%%%%%%%%%%%%%
Now one can define
%%%%%%%%%%%%%%%%%%%%%%%%%%%%%%%%%%%%%
\begin{align}
 \label{eq:als_b}
 [x_r^{(\kappa)}]_{z_\kappa} = c_r [\mathbf{X}_{r}^{(\kappa)}]_{z_\kappa} \nonumber\\
 [b_r^{(\kappa)}]_{z_\kappa} = \sum_{Z^\kappa} H_Z \, [\Omega_r^{\kappa}]_{Z^{\kappa}}
\end{align}
%%%%%%%%%%%%%%%%%%%%%%%%%%%%%%%%%%%%%
to arrive at the following working equations for obtaining optimal coefficients and SPOs
of the $\kappa$-th FS-DOF
%%%%%%%%%%%%%%%%%%%%%%%%%%%%%%%%%%%%%
\begin{align}
 \label{eq:als_org}
 [b_r^{(\kappa)}]_{z_\kappa} = \sum_{r^{\prime}} \left( S_{r,r^{\prime}}^{(\kappa)}
 +\epsilon \delta_{r,r^{\prime}} \right) [x_{r^{\prime}}^{(\kappa)}]_{z_\kappa} .
\end{align}
%%%%%%%%%%%%%%%%%%%%%%%%%%%%%%%%%%%%%
The above equation is a linear equation with respect to $r$ and can be solved with
standard linear algebra tools. Note that in Eq.(~\ref{eq:als_org}), the solution for
one FS-DOF $\kappa$ depends on the solution of all other FS-DOFs via the single hole overlap matrix $\mathbf{S}^{(\kappa)}$. Therefore, one needs
to solve it iteratively from an initial guess tensor, which constitutes the aforementioned ALS method.
\par
In the most general case, a major bottleneck of the above procedure is Eq.~(\ref{eq:als_b}), for 
$b^{(\kappa)}$ contains a multidimensional sum over index $Z^\kappa$ (i.e. overall tensor
elements except one) which in general cannot be separated into products of the low dimensional
sums. If the dimensionality of the system becomes too large, these sums become very expensive and
quickly unaffordable, especially since they need to be done multiple
times in each ALS iteration.
Additionally, one needs to store the full tensor, which grows
astronomically with the number of dimensions.

To circumvent this problem one can replace
the exact sum (integration) with the Monte Carlo integration scheme. First, a sampling space
is prepared by various sampling strategies and then the sum (integration) is done on
the sampling space. This is the procedure followed in the Monte Carlo CPD (MCCPD)
representation of the potential energy surface~\cite{Sch20:024108}.
Here, the smoothness of the potential energy surface is exploited in the sense that close grid points have similar energy cuts through the potential.
%with close reference points have a similar shape.

The generation of a sampling
 that captures the essence of the SQR electronic Hamiltonian tensor is, however, not
practicable, if not impossible. The reason is twofold: (i) The Hamiltonian tensor is very sparse. A sampling must be able to capture both, the zero and non-zero tensor elements in the sampling space
and (ii) there is no smoothness relation between neighboring tensor elements such that their values are highly erratic along all dimensions of the tensor.
Neighboring rows and columns of the tensor do not, in general, have a similar form. 
Finding a suitable sampling in this multidimensional space is, hence, a formidable
task.

%%%%%%%%%%%%%%%%%%%%%%%%%%%%%%%%%%%%%%%%%%%%%%%%%%%%%%%%%%%%%%%%%%%%%%%%%%%%%%%%%

\subsubsection{Sum-of-products to sum-of-products for CPD-SQR}

For practical purposes, we rely on the fact that the original
SQR Hamiltonian is already in a sum-of-products form. This is the Hamiltonian
in Eq.~(\ref{eq:ham_el_jw}).
This Hamiltonian has potentially a very large number of terms
as discussed in Section \ref{sec:theory:sop}, which makes its direct application
very costly. It can, however, be compressed into smaller a CPD-SQR Hamiltonian.
Assuming that the original Hamiltonian has already the desired sum-of-products form,
the summation over $Z^\kappa$, Eq.~(\ref{eq:als_b}), factorizes into sums of products of summations over $z_\kappa$, which renders the summation not only numerically affordable but also parallelizable.
Hence, the purpose of the CPD-SQR procedure outlined below is to find a much
more compact sum-of-products form than the original Hamiltonian while maintaining control over the accuracy of the compressed operator.

%In this section, we discuss a practicable approach to obtain the CPD-SQR form
%(Eq.~\ref{eq:candecomp}) of the Hamiltonian. The main disadvantage
%of the alternating least square approach described in Eq.~(\ref{eq:als_org}) is the
%construction of $b^{(\kappa)}$, i.e., the right-hand side of Eq~(\ref{eq:als_b})
%contains a sum over all but one primitive indices of the original tensor, which is
%generally non-separable.

Therefore, first, we discuss how to generate an
exact sum-of-product form of the SQR Hamiltonian that can be
used as a starting point for the compactification procedure.
In the FS-DOF primitive basis, all terms of Eq.~(\ref{eq:ham_el_jw})
that act only within the spin orbitals of one FS-DOF can be
summed up to form an uncorrelated operator term. Similarly,
one can form correlated operators (products acting on two or more
FS-DOF) by summing all terms that act within one FS-DOF
for each distinct string operating on the spin orbitals of
the other FS-DOF. The compact form of the SQR Hamiltonian for an
arbitrary FS-DOF combination is given in Ref.~\citenum{Sas22:134102} and
called the summed SQR (S-SQR) Hamiltonian. The S-SQR form of the
Hamiltonian is exact within the space of configurations spanned by the different FS-DOFs.
Since the matrix operators acting on each FS-DOF and
for different products are not related to each other, this
form of the SQR operator is \emph{de facto} an exact CPD:
%%%%%%%%%%%%%%%%%%%%%%%%%%%%%%%%%%%%%
\begin{align}
    \label{eq:exact_cpd}
    H_{z_1,\ldots,z_f} = \sum_{s=1}^{R_e} \prod_{\kappa=1}^f [\mathbf{Y}_s^{(\kappa)}]_{z_\kappa}.
\end{align}
%%%%%%%%%%%%%%%%%%%%%%%%%%%%%%%%%%%%%
Once, the SQR Hamiltonian is written as an exact SOP
form (Eq.~\ref{eq:exact_cpd}), Eq.(~\ref{eq:als_b}) can be written by means of
separable terms, to be exploited in numerical calculations
%%%%%%%%%%%%%%%%%%%%%%%%%%%%%%%%%%%%%
\begin{align}
 [b_r^{(\kappa)}]_{z_\kappa} &= \sum_{Z^\kappa} H_Z [\Omega_r^{\kappa}]_{Z^{\kappa}} \nonumber\\
 &= \sum_{r^{\prime}}^{R_e} 
[\mathbf{Y}_{r^{\prime}}^{(\kappa)}]_{z_{\kappa}}
 \prod_{\kappa^{\prime} \ne \kappa} \sum_{z_{\kappa'}}
[\mathbf{Y}_{r^{\prime}}^{(\kappa^{\prime})}]_{z_{\kappa^{\prime}}}
[\mathbf{X}_r^{(\kappa^{\prime})}]_{z_{\kappa^{\prime}}}.
 \label{eq:b_sop}
\end{align}
%%%%%%%%%%%%%%%%%%%%%%%%%%%%%%%%%%%%%

With Eq.~(\ref{eq:b_sop}) one may now straightforwardly implement the ALS algorithm and obtain a more compact SOP form of the Hamiltonian. At this stage, however, the resulting CPD-SQR Hamiltonian is
not guaranteed to be Hermitian. This is even true if all SPO $\mathbf{X}_r^{(\kappa)}$ of the initial guess tensor are chosen Hermitian  (apart from special cases where all operators $b_r^{(\kappa)}$ are Hermitian).
A simple workaround is the \emph{a posteriory} symmetrization of the  CPD-SQR Hamiltonian.
This, however, doubles the number of terms.
Hence, we use a symmetrized ansatz for the CPD-SQR Hamiltonian and adapt the ALS algorithm
to preserve Hermiticity during the optimization. We assume 
\begin{align}
H^{\text{CPD}} & = \sum_r^{R/2}  c_r  \prod_\kappa \mathbf{X}_r^{(\kappa)} +   \sum_r^{R/2} c_r^* \prod_\kappa\left(\mathbf{X}_r^{(\kappa)}\right)^\dagger  \label{eq:hcp:sym}\\
 & = \sum_r^{R}  \tilde c_r  \tilde {\mathbf{X}}_r^{(\kappa)},\label{eq:hcp:sym2}
\end{align}
i.e., we split the total sum into two parts, one part containing operators and coefficients to be found and a second part being the Hermitian conjugate of the first. The sum in the second line, Eq.~(\ref{eq:hcp:sym2}) simply runs over both sums in Eq.~(\ref{eq:hcp:sym}) where we assume that the Hermitian conjugate terms have indices $r=R/2+1, \cdots, R$.

Note that Hermitian conjugation of the complete $H^{\rm CPD}$, Eq.~(\ref{eq:hcp:sym2}), is equivalent to swapping the positions of the terms $r$ and $r+R/2$, that is, the Hermitian conjugate terms now have indices $r=1, \cdots, R/2$, and the original terms have indices $r>R/2$. The result of the sum is the same, of course. 

If the two different orders of the summation (with and without Hermitian conjugation) are inserted into Eq. (\ref{eq:jcp}) and following, the single-hole overlaps (assuming all operators and all $\tilde c_r$ are real) obey the permutation rules
\begin{align}
S^{(\kappa)}_{r,r'} & = S^{(\kappa)}_{r+R/2,r'+R/2}, \label{eq:s:sym}\\
S^{(\kappa)}_{r,r'+R/2}& = S^{(\kappa)}_{r+R/2,r'}.  
\end{align}
If Eq.\ \ref{eq:hcp:sym} should hold, one must also obtain
\begin{align}
b_r^{(\kappa)} = \left(b_{r+R/2}^{(\kappa)}\right)^\dagger .
\label{eq:b:sym}
\end{align}
The latter permutation rule, Eq. (\ref{eq:b:sym}), is usually not fulfilled when calculating $b_r^{(\kappa)}$ in Eq. (\ref{eq:als_b}), even when starting from a Hermitian initial guess. We hence enforce Eq. (\ref{eq:b:sym}) by symmetrizing the quantities  $S^{(\kappa)}$ and $b^{(\kappa)}$ in Eq.\ (\ref{eq:hcp:sym}) in each iteration step according to Eqs.\ (\ref{eq:s:sym}) to (\ref{eq:b:sym}). Thereafter, solving Eq. (\ref{eq:als_org}) with symmetrized $S^{(\kappa)}$ and $b^{(\kappa)}$ 
leads to a Hermitian CPD-SQR Hamiltonian in the form of Eq. (\ref {eq:hcp:sym}).  Note that, differently from an {\it a posteriory} Hermitization of the optimized Hamiltonian $H^{\rm CPD}$ according to Eq. (\ref{eq:hcp:sym}), here both, the original and the Hermitian conjugate SPOs, enter the working equations such that the optimization result is adapted to the ansatz Eq. (\ref{eq:hcp:sym}).

We implemented the scheme outlined above, which proved to be working but is numerically expensive as relatively large matrices need to be stored and processed.
Especially calculating the single-hole overlap matrices $S^{(\kappa)}$ is expensive, even though many temporary results, for instance, Eq. (\ref{eq:zeta}), can be re-used for several modes. Hence, we pick up the idea of finding a minimal but (almost) optimal and orthogonal basis of elementary operators $\mathbf{\Gamma}^{(\kappa)}_{\alpha_\kappa}$ such that any single particle operator of the original Hamiltonian can be expressed as a liner combination of elementary operators as 
\begin{align}
[\mathbf{Y}_{r'}^{(\kappa)}]_{z_{\kappa}} = \sum_{\alpha_\kappa}^{g_\kappa} a_{r',\alpha_\kappa} [\mathbf{\Gamma}^{(\kappa)}_{\alpha_\kappa}]_{z_{\kappa}}
\label{eq:spo_expansion}
\end{align}
with the expansion coefficients 
\begin{align}
a_{r,\alpha_\kappa} = \sum_{z_{\kappa}}
[\mathbf{\Gamma}^{(\kappa)}_{\alpha_\kappa}]_{z_{\kappa}}
[\mathbf{Y}_r^{(\kappa)}]_{z_{\kappa}}    
\end{align}
and expansion order $g_\kappa$. 
Of course, the SPO $\mathbf{X}_r^{(\kappa)}$ of $H^{\rm CPD}$ are also linear combinations of the same basis operators
\begin{align}
[\mathbf{X}_{r}^{(\kappa)}]_{z_{\kappa}} = \sum_{\alpha_\kappa}^{g_\kappa} c_{r,\alpha_\kappa} [\mathbf{\Gamma}^{(\kappa)}_{\alpha_\kappa}]_{z_{\kappa}}
\label{eq:x:svd}
\end{align}
as well as the single-hole overlaps of $H^{\rm CPD}$ with the original Hamiltonian
\begin{align}
[b_{r}^{(\kappa)}]_{z_{\kappa}} = \sum_{\alpha_\kappa}^{g_\kappa} d_{r,\alpha_\kappa} [\mathbf{\Gamma}^{(\kappa)}_{\alpha_\kappa}]_{z_{\kappa}}.
\end{align}
Here one hopes that the expansion order is much smaller than the number of matrix elements in the SPOs such that the overlaps  in Eq. (\ref{eq:zeta}) can be expressed as
\begin{align}
 \zeta_{r,r^{\prime}}^{(\kappa^{\prime})} & = \sum_{z_\kappa} [\mathbf{X}_r^{(\kappa)}]_{z_\kappa} [\mathbf{X}_{r^{\prime}}^{(\kappa)}]_{z_\kappa} \\
& = \sum_{\alpha_\kappa}^{g_\kappa} c_{r,\alpha_\kappa}c_{r',\alpha_\kappa}
\end{align}
which are dot-products of small vectors as opposed to large matrices. Similarly, Eq. (\ref{eq:b_sop}) becomes after multiplication with $\mathbf{\Gamma}^{(\kappa)}_{\alpha_\kappa}$ from the left and subsequent summation over $z_\kappa$
\begin{align}
 d_{r,\alpha_\kappa}
 &= \sum_{r^{\prime}}^{R_e} 
 a_{r',\alpha_\kappa} 
 \prod_{\kappa^{\prime} \ne \kappa} 
 \sum_{\alpha_\kappa}^{g_\kappa}
 a_{r',\alpha_{\kappa'}} c_{r,\alpha_{\kappa'}}, 
 \label{eq:b_sop:svd}
\end{align}
again only involving small dot-products. 

In practice, the basis operators $\Gamma^{(\kappa)}_{\alpha_\kappa}$ are not obtained via diagonalization of the density matrix, Eq. (\ref{eq:density:tucker}), as done for the Tucker decomposition, because calculating the density matrix is numerically expensive, but by re-shaping all $\mathbf{Y}_{r'}^{(\kappa)}$ into vectors and arranging them into columns of a possibly large matrix $U^{(\kappa)}$ of size $\mathcal{N}^2_\kappa \times R_e$. Subsequently the $g_\kappa$ most important left singular vectors  and values $\sigma_{\alpha_\kappa}$ are computed, where $g_\kappa$ is determined by a pre-set cut-off $\epsilon_{\rm SVD}$ (typically set to $10^{-9}$) such that 
\begin{align}
  \frac{ \sum\limits_{\alpha_\kappa = g_\kappa + 1}^{\mathcal{N}_\kappa^2} \sigma_{\alpha_\kappa}^2 }{\sum\limits_{\alpha_\kappa =  1}^{\mathcal{N}_\kappa^2} \sigma_{\alpha_\kappa}^2}
  < \epsilon_{\rm SVD}.
\end{align}
The respective left singular vectors are reshaped into the matrices $\mathbf{\Gamma}^{(\kappa)}_{\alpha_\kappa}$ and serve as basis operators. This has to be done only once at the beginning of the calculation. 
The ALS optimization is hence performed in a much smaller vector space than the original SPOs are defined in, which saves net computational effort. Only at the very end of the calculation, the operators $\mathbf{X}_{r}^{(\kappa)}$ are reconstructed using Eq. (\ref{eq:x:svd}). 

A major drawback of this procedure is, however, that Hermiticity cannot be restored with the recipe outlined above. Here we resort to a different strategy. As in all of the formulas above, we assume that the matrix elements $[\mathbf{Y}_{r'}^{(\kappa)}]_{z_\kappa}$ of the original SPOs are real. We can then decompose all SPOs into a symmetric (Hermitian) part $ ^{\rm S}\mathbf{Y}_{r'}^{(\kappa)}$ and an anti-symmetric (anti-Hermitian) part $ ^{\rm A}\mathbf{Y}_{r'}^{(\kappa)}$
\begin{align}
    \mathbf{Y}_{r'}^{(\kappa)} = {}^{\rm S} \mathbf{Y}_{r'}^{(\kappa)} + {}^{\rm A} \mathbf{Y}_{r'}^{(\kappa)}
\end{align}
Furthermore, one may realize that 
\begin{align}
  \sum_{z_\kappa} \left[ {}^{\rm S}\mathbf{Y}_{r'}^{(\kappa)} \right]_{z_\kappa}
  \cdot \left[ {}^{\rm A}\mathbf{Y}_{r'}^{(\kappa)} \right]_{z_\kappa} = 0
\end{align}
for any symmetric and anti-symmetric matrices. We can then repeat the procedure to obtain elementary operators as outlined above, but separately for the symmetric and the anti-symmetric operators, i.e, all symmetric matrices  $ ^{\rm S}\mathbf{Y}_{r'}^{(\kappa)}$ are reshaped into vectors and arranged into columns of a matrix $ ^{\rm S}\mathbf{U}_{r'}^{(\kappa)}$ and all anti-symmetric matrices  $ ^{\rm A}\mathbf{Y}_{r'}^{(\kappa)}$ are reshaped into vectors and arranged into columns of a matrix $ ^{\rm A}\mathbf{U}_{r'}^{(\kappa)}$  from which  $^{\rm S}g_\kappa$ symmetric and $^{\rm A}g_\kappa$ anti-symmetric left singular vectors (elementary operators) $^{\rm S} \mathbf{\Gamma}^{(\kappa)}_{\alpha_\kappa}$ and $^{\rm A} \mathbf{\Gamma}^{(\kappa)}_{\beta_\kappa}$ are computed, respectively. The combined set, $\left\{ {}^{\rm S} \mathbf{\Gamma}a^{(\kappa)}_{\alpha_\kappa}, {}^{\rm A} \mathbf{\Gamma}^{(\kappa)}_{\beta_\kappa}\right\}$ hence forms an orthonormal basis and such that 
\begin{align}
[\mathbf{Y}_{r'}^{(\kappa)}]_{z_{\kappa}} & = \sum_{\alpha_\kappa=1}^{{}^{\rm S}g_\kappa} {}^{\rm S}a_{r',\alpha_\kappa} [^{\rm S}\mathbf{\Gamma}^{(\kappa)}_{\alpha_\kappa}]_{z_{\kappa}} + 
\sum_{\beta_\kappa=1}^{{}^{\rm A}g_\kappa} {}^{\rm A}a_{r',\beta_\kappa} [^{\rm A}\mathbf{\Gamma}^{(\kappa)}_{\beta_\kappa}]_{z_{\kappa}} \\
& =  \sum_{\gamma_\kappa=1}^{\bar g_\kappa={}^{\rm S}g_\kappa + {}^{\rm A}g_\kappa} \bar a_{r',\gamma_\kappa} [ \mathbf{\bar \Gamma}^{(\kappa)}_{\gamma_\kappa}]_{z_{\kappa}}
\label{eq:y:symexpand}
\end{align}
with expansion coefficients ${}^{\rm S}a_{r',\alpha_\kappa}$ and ${}^{\rm A}a_{r',\beta_\kappa}$ for the symmetric and anti-symmetric elementary operators respectively (or combined expansion coefficients $ \bar a_{r',\gamma_\kappa}$ for the combined set of basis operators). 
Similar formulas hold for $\mathbf{X}_{r}^{(\kappa)}$ and $b_{r}^{(\kappa)}$. The Hermitian conjugate of the expansion, Eq. (\ref{eq:y:symexpand}) is obtained by simply swapping the sign of the coefficients ${}^{\rm A}a_{r',\beta_\kappa}$ of the anti-symmetric basis operators. 

We are now in a position to return to the ansatz Eq. (\ref{eq:hcp:sym}) and repeat the derivations, Eqs. (\ref{eq:b:sym}) to (\ref{eq:b_sop:svd}). One realizes that the optimization procedure can again be performed entirely in the vector space of the expansion coefficients and the symmetrization of Eq. (\ref{eq:b:sym}) is simply performed by swapping the sign of the coefficients of the anti-symmetric basis operators.  

Note furthermore, that only non-zero vectors  $ ^{\rm S}\mathbf{Y}_{r'}^{(\kappa)}$ and  $ ^{\rm A}\mathbf{Y}_{r'}^{(\kappa)}$ need to be stored in the matrices  $ ^{\rm S}\mathbf{U}_{r'}^{(\kappa)}$ and  $ ^{\rm A}\mathbf{U}_{r'}^{(\kappa)}$, respectively. If all $ ^{\rm A}\mathbf{Y}_{r'}^{(\kappa)} = 0$ (i.e., all SPO of the original Hamiltonian are Hermitian) the fit $H^{\rm CPD}$ is obtained Hermitian automatically without Ansatz Eq. (\ref{eq:hcp:sym}) irrespective of whether the reduction, Eq. (\ref{eq:spo_expansion}) is used. 

We have implemented the SOP-to-SOP algorithms outlined above in a new program {\tt compactoper} as part of the Heidelberg MCTDH package~\cite{mctdh:MLpackage}.
An important comment related to its general application to 
nuclear dynamics problems should be added here:
While the procedure outlined above works in general for any operators in
sum-of-products form in the sense that it leads to (much) fewer
terms in the sum, the resulting operator may turn out not to be the optimal
form to perform dynamics calculations.
For instance, kinetic energy operators in polyspherical coordinates\cite{Cha92:6217,Gat09:1} may,
depending on the molecule under consideration,
contain thousands of product terms.\cite{sch22:6170} Many of the single particle operators are, however, unit operators or diagonal operators that depend on a single coordinate and require none or little numerical effort to operate on the wave function.
The procedure above, however, will return all single particle operators in matrix form, which are expensive to operate.
Operating the kinetic energy in its original form would usually by far outperform operating its compacted form obtained with the schemes outlined above.
One should, hence, use this method only if the vast majority
of SPOs of the original Hamiltonian are matrices in the first place,
as is the case for the SQR calculations shown below,
or if all SPOs are diagonal in the first place,
i.e., if the Hamiltonian is a potential
(in this case, our implementation uses vectors of single-particle
potentials instead of matrices of single-particle operators).

%%%%%%%%%%%%%%%%%%%%%%%%%%%%%%%%%%%%%%%%%%%%%%%%%%%%%%%%%%%%%%%%%%%%%%%%
%%%%%%%%%%%%%%%%%%%%%%%%%%%%%%%%%%%%%%%%%%%%%%%%%%%%%%%%%%%%%%%%%%%%%%%%

%%%%%%%%%%%%%%%%%%%%%%%%%%%%%%%%%%%%%%%%%%%%%%%%%%%%%%%%%%%%%%%%%%%%%%%%
\section{Numerical examples} \label{sec:results}
The \emph{ab initio} MCTDH-SQR approach can be directly
applied to the calculation of electronic 
excitation or ionization spectra of molecular systems
through time propagation instead of matrix diagonalization. For this, one needs
to prepare an appropriate initial state (ionized and excited in cases of ionization and
excitation spectrum, respectively) that overlaps with the desired state for
time-propagation. Then, the power spectrum can be obtained as usual from the
Fourier transform of the autocorrelation function
%%%%%%%%%%%%%%%%%%%%%%%%%%%%%%%%%%%%%
\begin{align}
 \sigma(E) = \frac{1}{\pi}Re\int_{0}^{\infty} e^{iEt}
 \langle \Psi|\Psi(t) \rangle dt .
 \label{eq:power_spectrum}
\end{align}
%%%%%%%%%%%%%%%%%%%%%%%%%%%%%%%%%%%%%
Here we benchmark the method on water, trans-polyene, and
glycine.

\subsection{H$_2$O} \label{sec:res:h2o}
\subsubsection{Ionization}

We compare the compactness of the T-SQR and CPD-SQR Hamiltonian first by
considering the ionization spectrum of the H$_2$O molecule as a benchmark.
The molecular orbitals (MO) are generated using the
6-31G atomic basis for both O and H, and
the lowest energy MO (1s of O) is frozen.
The remaining 12 spatial orbitals are
considered for the calculation.
%
%Three FS-DOFs are formed each consisting of 4 spatial (8 spin) orbitals.
%
%We prune the configuration space in each FS-DOF in the following way:
%
The scheme of FS-DOFs and the corresponding active spaces are given in
Table~\ref{tab:h2o_fsdof_ip}.
%%%%%%%%%%%%%%%%%%%%%%%%%
% Make Table
%% Table: H2O: FS-DOF
\begin{table}[t]
 \caption{FS-DOFs and active spaces for the ionization spectrum calculation of H$_2$O.
          }
 \begin{ruledtabular}
  {\begin{center}
    \begin{tabular}{lccccr}
    FS-DOF & No. spat. orb. & \multicolumn{3}{c}{Occupations} & No. conf.\\
    \cline{3-5}
           &              & alpha & beta & Total         &  \\
     \hline
       I   & 4  & 2-4 & 2-4 & 6-8 & 37 \\
       II  & 4  & 0-2 & 0-2 & 0-2 & 37 \\
       III & 4  & 0-2 & 0-2 & 0-2 & 37
    \end{tabular}
   \end{center} }
  \end{ruledtabular}
 \label{tab:h2o_fsdof_ip} 
\end{table}
%%%%%%%%%%%%%%%%%%%%%%%%%
% (i) In the first FS-DOF, we allow (2$-$4) alpha, (2$-$4) beta, and (6$-$8) total electrons.
%
% (ii) In the second FS-DOF, we allow (0$-$2) alpha, (0$-$2) beta, and (0$-$2) total electrons
% and (iii) in the third FS-DOF, we allow (0$-$2) alpha, (0$-$2) beta and (0$-$2) total electrons.
% This generates 37 configurations in each of the three FS-DOFs.
%%%%%%%%%%%%%%%%%%%%%%%%%
The atomic basis and FS-DOFs are the same as in Ref.~\citenum{Sas22:134102},
where we compiled the S-SQR and T-SQR results.
The total number of terms in the original SQR
Hamiltonian is 40618, which goes down to
6890 when summed up without approximations, namely
the S-SQR (exact CPD).

The initial wavefunction for the propagation
is generated by applying the
ionization operator
%%%%%%%%%%%%%%%%%%%%%%%%%%%%%%%%%%%%%
\begin{align}
 \label{eq:iOph2o}
 \hat{A} = \sum_{i=1}^{4}\sum_{\sigma=\uparrow, \downarrow}
           \hat{a}_{i,\sigma}
\end{align}
%%%%%%%%%%%%%%%%%%%%%%%%%%%%%%%%%%%%%
to the ground electronic state of neutral H$_2$O.
This initial wavefunction is a spin doublet and a linear combination
of singly-ionized configurations.
%%%%%%%%%%%%%%%%%%%%%%%%%%%%%%%%%%%%%
\begin{table}[t] 
 \caption{Number of Hamiltonian terms and wall-clock time for the different
          sum-of-product form of the Hamiltonian.
          The wall-clock time is given for 20 fs time propagation.
          The calculations have been performed with 16 CPUs using shared-memory
          parallelization on the same machine and CPU type,  namely,  Dual-Core
          Intel Xeon,  processor type E5-2650 v2 running at 2.6 GHz and the
          wall-clock times are intended for their relative comparison only.
          The SQR, S-SQR, and T-SQR results are reproduced from
          J. Chem. Phys. {\bf 157}, 134102 (2022) [Ref.~\citenum{Sas22:134102}].
          }
 \begin{ruledtabular}
  {\begin{center}
    \begin{tabular}{lccr}
      Method & Hamil. terms & Time (h:m)\\
      \hline
      SQR     & 40618  & 243:08 \\
      S-SQR   & 6890   & 58:21 \\
      T-SQR   & 45000  & 404:29 \\
      \hline
      CPD-SQR
              & 200    & 2.08 \\
              & 400    & 3.33 \\
              & 500    & 4.08 \\
              & 600    & 4.48
     \end{tabular}
  \end{center} }
 \end{ruledtabular}
 \label{tab:h2o_ip} 
\end{table}
%%%%%%%%%%%%%%%%%%%%%%%%%%%%%%%%%%%%%
%%%%%%%%%%%%%%%%%%%%%%%%%%%%%%%%%%%%%
\begin{figure}[t]
\begin{center}
\includegraphics[width=9.0cm]{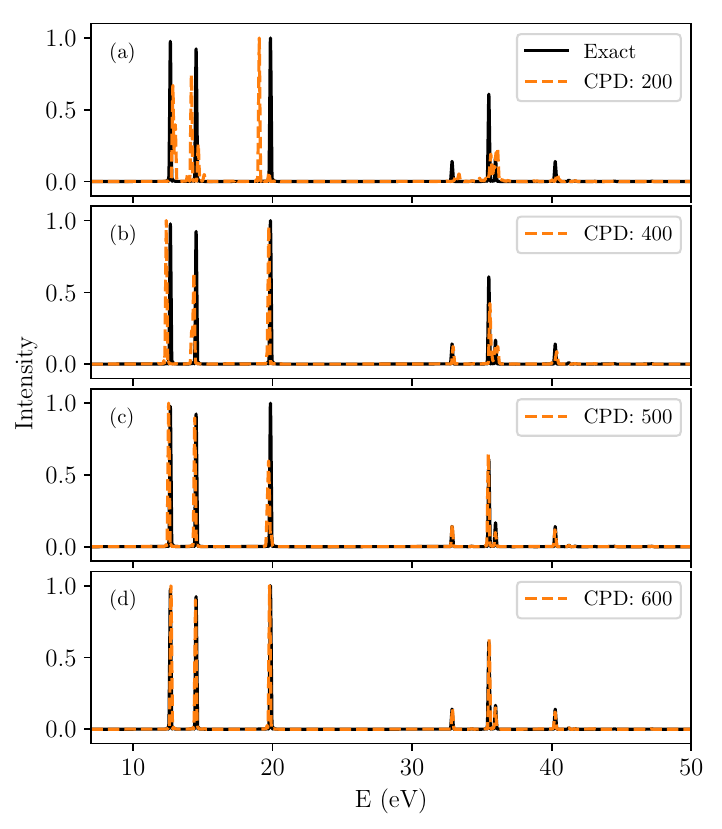}
\includegraphics[width=9.0cm]{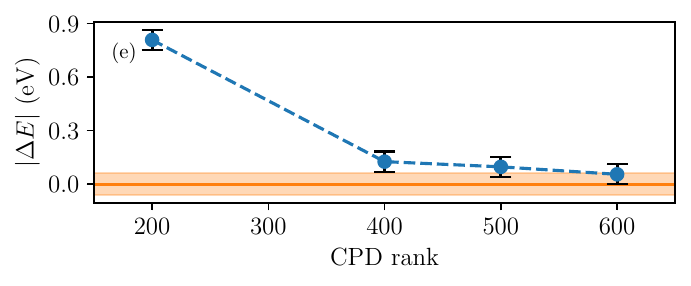}
\caption{Convergence of electronic ionization spectra of H$_2$O
         with different CPD rank. The spectrum in black is calculated with
         the exact Hamiltonian. The spectra in red are calculated with CPD-SQR
         Hamiltonian with 200, 400, 500, and 600 CPD ranks and shown in
         (a), (b), (c), and (d), respectively.
         The convergence of the 3rd peak (centered
         around 20 eV) with respect to the CPD rank is shown in (e).
         The error bar represents the spectral resolution, full width at half maximum (FWHM),
         stemming from the propagation time.
         The autocorrelation function was propagated for 20 fs, resulting in a FWHM of about 0.1 eV.
         }
\label{fig:h2o_ip}
\end{center}
\end{figure}
%%%%%%%%%%%%%%%%%%%%%%%%%%%%%%%%%%%%%
Fig.~\ref{fig:h2o_ip} compares the electronic ionization spectrum of H$_2$O using
the exact Hamiltonian with CPD-SQR Hamiltonians of different CPD ranks.
As expected, the numerical accuracy of the spectrum increases with increasing
CPD terms and the convergence is achieved with only $\sim$ 600 Hamiltonian terms.
Table~\ref{tab:h2o_ip} compares the number of Hamiltonian terms and CPU time
for different SOP representations of the electronic Hamiltonian.
%
%The SQR, S-SQR, and T-SQR results are taken from Ref.~\cite{Sas22:134102}.
As seen by inspecting Table~\ref{tab:h2o_ip}, the CPD-SQR form significantly
outperforms T-SQR in achieving a more compact SOP form of the electronic
Hamiltonian. In this example, the CPD-SQR compression factors
are $\mathcal{R}=68$ and $\mathcal{R}=12$
with respect to the original SQR and S-SQR Hamiltonian, respectively,
where $\mathcal{R}$ is defined as
the number of terms of the exact Hamiltonian divided by the number of terms 
of the corresponding compressed Hamiltonian.
%%%%%%%%%%%%%%%%%%%%%%%%%%%%%%%%%%%%%
%\begin{align}
% \label{eq:comp_fac}
% \mathcal{R} = \frac{\text{Number of terms of the original Hamiltonian}}{\text{CPD-SQR terms to achieve numerical %convergence}}
%\end{align}
%%%%%%%%%%%%%%%%%%%%%%%%%%%%%%%%%%%%%

\subsubsection{Excitation}
%%%%%%%%%%%%%%%%%%%%%%%%%%%%%%%%%%%%%
%% Table: H2O: FS-DOF
\begin{table}[t]
 \caption{FS-DOFs and active spaces for the excitation spectrum calculation of H$_2$O.
          }
 \begin{ruledtabular}
  {\begin{center}
    \begin{tabular}{lccccr}
    FS-DOF & No. spat. orb. & \multicolumn{3}{c}{Occupations} & No. conf.\\
    \cline{3-5}
           &              & alpha & beta & Total         &  \\
     \hline
       I   & 4  & 1-4 & 1-4 & 5-8 & 93 \\
       II  & 4  & 0-3 & 0-3 & 0-3 & 93 \\
       III & 5  & 0-2 & 0-2 & 0-2 & 56 \\
       IV  & 5  & 0-2 & 0-2 & 0-2 & 56 \\
       V   & 5  & 0-2 & 0-2 & 0-2 & 56 \\
    \end{tabular}
   \end{center} }
  \end{ruledtabular}
 \label{tab:h2o_fsdof} 
\end{table}
%%%%%%%%%%%%%%%%%%%%%%%%%%%%%%%%%%%%%
%%%%%%%%%%%%%%%%%%%%%%%%%%%%%%%%%%%%%
\begin{figure}[t]
\begin{center}
\includegraphics[width=9.0cm]{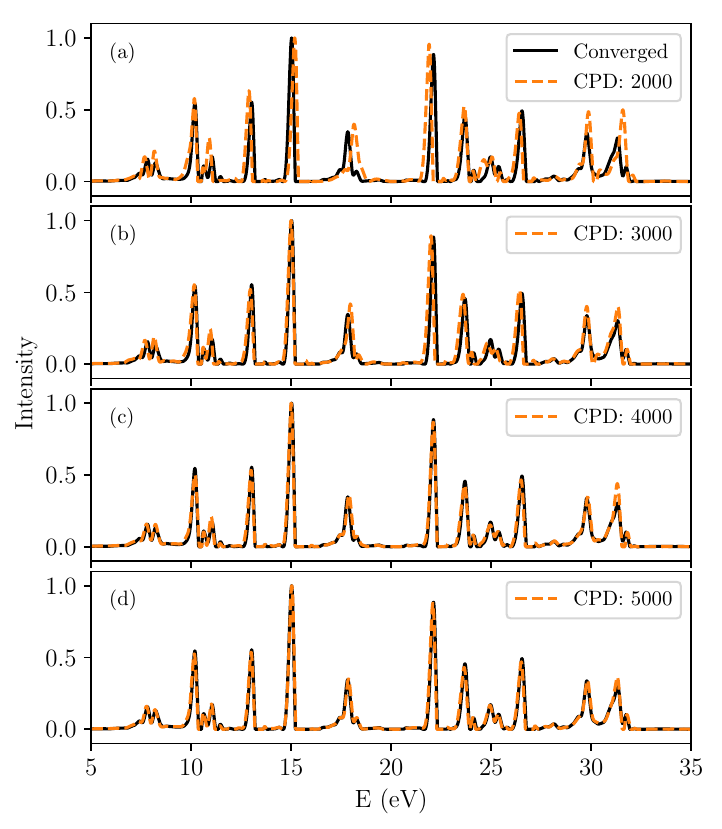}
\includegraphics[width=9.0cm]{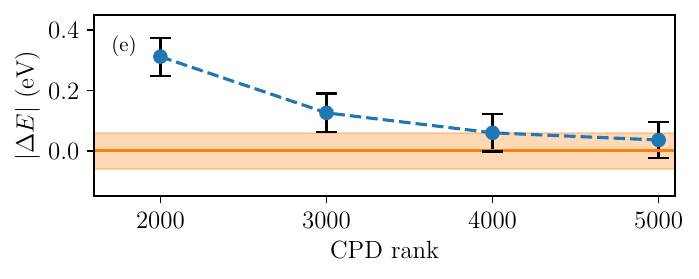}
\caption{Convergence of electronic excitation spectra of H$_2$O
         with different CPD ranks.
         The spectra in red are calculated with CPD-SQR
         Hamiltonian with 2000, 3000, 4000, and 5000 CPD rank and shown in
         (a), (b), (c), and (d), respectively.
         The numerically converged (regarding Hamiltonian terms) spectra are shown in black and calculated with
         10000 CPD rank. Further increases in the CPD rank do not change the spectrum.
         (e) Convergence of the peak centered around 18 eV with respect to
         the CPD rank.
         For details see the caption of Fig~\ref{fig:h2o_ip}.
         }
\label{fig:h2o_ex}
\end{center}
\end{figure}
%%%%%%%%%%%%%%%%%%%%%%%%%%%%%%%%%%%%%
Similarly to ionization, the positions of the
low-lying excited states of H$_2$O can be calculated
by time-propagation.
In this example, we generate MOs using the cc-pVDZ basis for H and O.
The lowest energy MO (1s of O) is frozen and the remaining 46 spin orbitals
are considered for the calculation.
The scheme of FS-DOFs and the corresponding active spaces are given in
Table~\ref{tab:h2o_fsdof}.
The initial wavefunction for the propagation is generated by applying the excitation
operator 
%%%%%%%%%%%%%%%%%%%%%%%%%%%%%%%%%%%%%
\begin{align}
 \label{eq:eOpch2o}
 \hat{E} = \sum_{i=2}^{4}\sum_{a=5}^{8}\sum_{\sigma=\uparrow, \downarrow}
           \hat{a}_{a,\sigma}^{\dagger}\hat{a}_{i,\sigma} .
\end{align}
%%%%%%%%%%%%%%%%%%%%%%%%%%%%%%%%%%%%%
on the ground electronic state of H$_2$O.
The initial wavefunction is a spin-singlet superposition of singly excited
configurations and overlaps with many of the low-lying excited
singlet states of H$_2$O.
Note that $\hat{E}$ is not the dipole operator and therefore the corresponding
excitation spectrum does not correspond to an absorption spectrum, which would
contain much fewer peaks.
The total number of terms in the original SQR
Hamiltonian is 544364 and in the S-SQR Hamiltonian before compression
already goes down to 63119 terms.
Fig.~\ref{fig:h2o_ex} compares the electronic excitation spectrum of H$_2$O
using CPD-SQR Hamiltonians with increasing rank.
The spectrum in black is by far numerically converged with CPD rank 10000.
From the figure, it is clear that the convergence is achieved already with
about 4000 CPD-SQR Hamiltonian terms.
The CPD-SQR compression factor is $\mathcal{R}=136$
and $\mathcal{R}=16$ 
with respect to the original SQR and S-SQR Hamiltonian, respectively.

Considering now the ML-MCTDH-SQR representation of the electronic wavefunction,
the primitive space corresponds to $1.5\cdot 10^{9}$ configurations (product of
the rightmost column in Table~\ref{tab:h2o_fsdof}).
%These can be compared to the $78\cdot 10^{6}$ configurations in the Hilbert space of 8 electrons in 23 spatial orbitals.
These can be compared to the $1.6\cdot 10^{5}$ configurations in the Hilbert space of 8 electrons in
the restricted Fock space generated by the FS-DOF formation in Table~\ref{tab:h2o_fsdof}.
Of course, the ML-MCTDH-SQR ansatz spans a primitive Fock space much
larger than the Hilbert space of the relevant electronic number.
However, the number of propagated wavefunction coefficients is $\sim 2\cdot 10^{4}$
which is significantly smaller than the number of CI coefficients.
%
%\emph{Oriol: Add number here, hopefully significantly smaller than the number of full-CI coefficients}.

%%%%%%%%%%%%%%%%%%%%%%%%%%%%%%%%%%%%%%%%%%%%%%%%%%%%%%%%%%%%%%%%%%%%%%%%
%%%%%%%%%%%%%%%%%%%%%%%%%%%%%%%%%%%%%%%%%%%%%%%%%%%%%%%%%%%%%%%%%%%%%%%%
\subsection{Trans-C$_8$H$_{10}$}  \label{sec:res:polyene}
%%%%%%%%%%%%%%%%%%%%%%%%%%%%%%%%%%%%%%%%%%%%%%%%%%%%%%%%%%%%%%%%%%%%%%%%
%%%%%%%%%%%%%%%%%%%%%%%%%%%%%%%%%%%%%
\begin{figure}[t]
\begin{center}
\includegraphics[width=9.0cm]{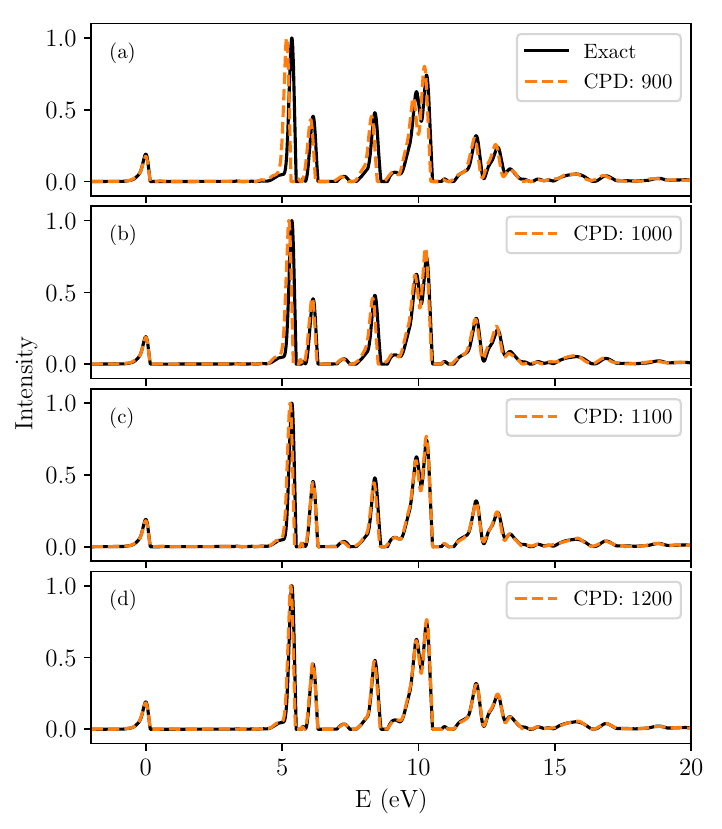}
\includegraphics[width=9.0cm]{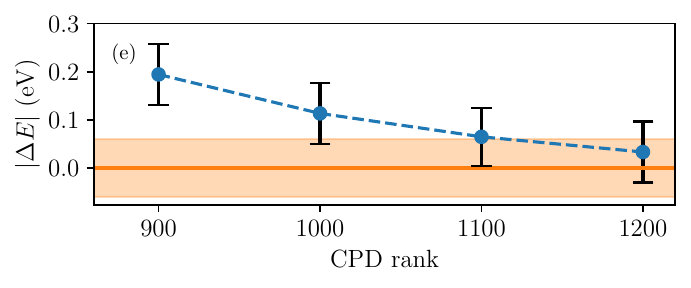}
\caption{Convergence of excitation spectra of trans-C$_8$H$_{10}$ with different
         CPD rank.
         The spectrum in black is calculated with
         the exact Hamiltonian. The spectra in red are calculated with CPD-SQR
         Hamiltonian with 900, 1000, 1100, and 1200 CPD ranks and shown in
         (a), (b), (c), and (d), respectively.
         (e) Convergence of the peak centered around 5 eV (1st excited state)
         with respect to the CPD rank.
         For details see the caption of Fig~\ref{fig:h2o_ip}.
         }
\label{fig:c8h10_ex}
\end{center}
\end{figure}
%%%%%%%%%%%%%%%%%%%%%%%%%%%%%%%%%%%%%

In the following, we obtain the low-lying $\pi$-excited states of trans-C$_8$H$_{10}$.
% analogously to the smaller trans-C$_4$H$_{6}$ polyene. 
%Again, the all-trans geometry of C$_8$H$_{10}$
%is optimized at the B3LYP/cc-pVDZ label of theory
The geometry of C$_8$H$_{10}$ is taken to be all trans and is optimized at the
B3LYP/cc-pVDZ label of theory
and the MOs are calculated using the cc-pVDZ basis for C and H.
%
%We consider 8 $\pi$ electrons in 16 $\pi$ spatial (32 spins) orbitals,
The orbital space is chosen to be $\pi$-doublet valence, i.e., 8 $\pi$ electrons in 16 $\pi$ spatial (32 spin) orbitals,
and the details of the FS-DOF partitioning and active spaces are given in
Table~\ref{tab:c8h10_fsdof}.
%BLI.
%%%%%%%%%%%%%%%%%%%%%%%%%
%% Table: C8H10: FS-DOF
\begin{table}[t]
 \caption{FS-DOFs and active spaces for the excitation spectrum calculation of trans-C$_8$H$_{10}$.
          }
 \begin{ruledtabular}
  {\begin{center}
    \begin{tabular}{lccccr}
    FS-DOF & No. spat. orb. & \multicolumn{3}{c}{Occupations} & No. conf.\\
    \cline{3-5}
           &              & alpha & beta & Total         &  \\
     \hline
       I   & 4  & 1-4 & 1-4 & 5-8 & 93 \\
       II  & 4  & 0-3 & 0-3 & 0-3 & 93 \\
       III & 4  & 0-2 & 0-2 & 0-2 & 37 \\
       IV  & 4  & 0-2 & 0-2 & 0-2 & 37 
    \end{tabular}
   \end{center} }
  \end{ruledtabular}
 \label{tab:c8h10_fsdof} 
\end{table}
%%%%%%%%%%%%%%%%%%%%%%%%%
%Four FS-DOFs are formed each consisting of 8 spin orbitals. The FS-DOFs are constructed
%following way: (i) in FS-DOF-I, we allow (1-4) alpha, (1-4) beta, and (5-8) total electrons, (ii) in FS-DOF-II, we allow (0-3) alpha, (0-3) beta and (0-3) total electrons, and (iii-iv) in FS-DOF-III and -IV, (0-2) alpha, (0-2) beta and (0-2) total electrons are allowed.
%This generates 93 configurations in FS-DOF-I and -II, and 37 configurations in FS-DOF-III and -IV, respectively.
The initially excited wavefunction is generated as a superposition of singly-excited
configurations covering all combinations of the orbitals HOMO-2 to HOMO and LUMO to LUMO+3.
%%%%%%%%%%%%%%%%%%%%%%%%%%%%%%%%%%%%%
%\begin{align}
% \label{eq:eOpc8h10}
% \hat{E} = \sum_{i=2}^{4}\sum_{a=5}^{8}\sum_{\sigma=\uparrow, \downarrow}
%           \hat{a}_{a,\sigma}^{\dagger}\hat{a}_{i,\sigma} .
%\end{align}
%%%%%%%%%%%%%%%%%%%%%%%%%%%%%%%%%%%%%
From a total of 123648 terms in the original SQR Hamiltonian, we arrive at
an S-SQR (exact CPD) with 14212 terms.
The excitation spectrum obtained with different CPD ranks is compared with the spectrum
calculated with exact Hamiltonian and shown in Fig~\ref{fig:c8h10_ex}.
The CPD-SQR reaches convergences with about 1100 Hamiltonian terms, whereby
calculations with only about 900 terms show a good qualitative agreement with
the converged excitation energies.
%The CPD-SQR compression factors ($\mathcal{R}$) in this numerical example is
%$\sim$ 112 and 13 with respect to the original SQR and S-SQR Hamiltonian, respectively.

%%%%%%%%%%%%%%%%%%%%%%%%%%%%%%%%%%%%%
\begin{figure}[t]
\begin{center}
\includegraphics[width=9.0cm]{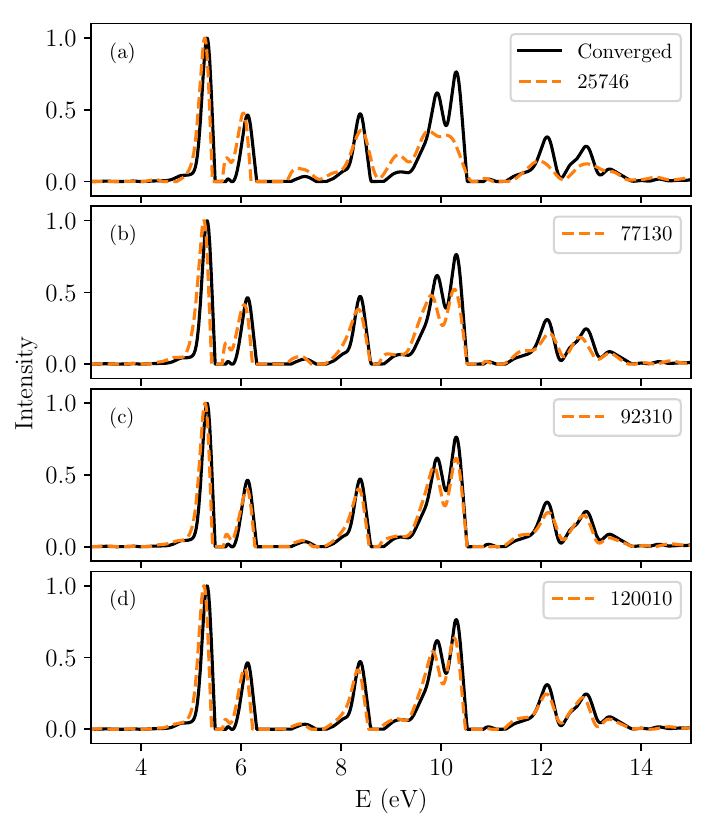}
\caption{Convergence of excitation spectra of trans-C$_8$H$_{10}$ with MCTDH coefficients.
         The spectrum in black is calculated with
         the converged MCTDH wavefunction where the lowest natural population of each
         node of the MCTDH wavefunction is less than $10^{-3}$.
         The spectra in red are calculated with 25746,
         77130, 92130, and 120010 MCTDH coefficients
         and shown in (a), (b), (c), and (d), respectively.
         }
\label{fig:c8h10_wfn}
\end{center}
\end{figure}
%%%%%%%%%%%%%%%%%%%%%%%%%%%%%%%%%%%%%
In this numerical example, the ML-MCTDH-SQR wavefunction consists of $1.2\cdot10^{7}$ primitive
configurations, of which only $4\cdot 10^{4}$ configurations correspond to a Hilbert space
of $4$ alpha and $4$ beta electrons. Although the ML-MCTDH wavefunction spans a significantly
larger primitive space than the pertinent Hilbert space, the number of propagated MCTDH
coefficients can be of the same order or less to get a qualitative agreement with the
converged spectrum. Fig.~\ref{fig:c8h10_wfn} compares the converged spectrum with
varying numbers of propagated MCTDH coefficients. The converged calculation is done
with MCTDH wavefunction where the lowest natural population of each node is less than
$10^{-3}$. It's clear that even with the MCTDH coefficients (25746) less than the
Hilbert space configuration ($4\cdot 10^{4}$), one obtains a qualitatively good agreement.
%(the peak position matches, only the intensity differs a bit.)

% \emph{Oriol: all this discussion about the polyenes reads quite boring and repetitive.
% Can we say something about the actual physics of these molecules? For example, if
% one generates a localized excitation on one side of the molecule (by properly choosing
% the excitation operator to generate the right linear combination), how does the excitonic
% dynamics proceed? Do we see some damping, as in glycine, due to the slow mixing up of other
% states into the picture? I know we had not discussed adding such aspects,
% but the text is quite dry and uninteresting as of now.
% Here and there, we should also say something about the wavefunction. We keep shifting
% this under the carpet but any clever referee will come at us with it.
%
% At least we should see
% what happens when the number of propagated coefficients falls below the number of
% coefficients of the corresponding Hilbert space for the number of orbitals and electrons.
% If we cannot beat this, then this is a very bad signal and we should think what to do to
% break this barrier, if possible.}
%%%%%%%%%%%%%%%%%%%%%%%%%%%%%%%%%%%%%%%%%%%%%%%%%%%%%%%%%%%%%%%%%%%%%%%%

%%%%%%%%%%%%%%%%%%%%%%%%%%%%%%%%%%%%%%%%%%%%%%%%%%%%%%%%%%%%%%%%%%%%%%%%
%%%%%%%%%%%%%%%%%%%%%%%%%%%%%%%%%%%%%%%%%%%%%%%%%%%%%%%%%%%%%%%%%%%%%%%%
\subsection{Glycine} \label{sec:res:glycine}
%%%%%%%%%%%%%%%%%%%%%%%%%%%%%%%%%%%%%%%%%%%%%%%%%%%%%%%%%%%%%%%%%%%%%%%%
%%%%%%%%%%%%%%%%%%%%%%%%%%%%%%%%%%%%%%%%%%%%%%%%%%%%%%%%%%%%%%%%%%%%%%%%
%%%%%%%%%%%%%%%%%%%%%%%%%%%%%%%%%%%%%
%% Table: H2O: FS-DOF
\begin{table}[t]
 \caption{Constructed FS-DOFs for the ionization calculation of glycine molecule
          }
 \begin{ruledtabular}
  {\begin{center}
    \begin{tabular}{lccccr}
    FS-DOF & Spatial orb. & \multicolumn{3}{c}{Electron} & Conf. No.\\
    \cline{3-5}
           &              & alpha & beta & Total         &  \\
     \hline
       I   & 3  & 1-3 & 1-3 & 4-6 & 22 \\
       II  & 3  & 1-3 & 1-3 & 4-6 & 22 \\
       III & 4  & 1-4 & 1-4 & 5-8 & 93 \\
       IV  & 4  & 0-3 & 0-3 & 0-3 & 93 \\
       V   & 4  & 0-2 & 0-2 & 0-2 & 37 \\
       VI  & 5  & 0-2 & 0-2 & 0-2 & 56 \\
    \end{tabular}
   \end{center} }
  \end{ruledtabular}
 \label{tab:glycine_fsdof} 
\end{table}
%%%%%%%%%%%%%%%%%%%%%%%%%%%%%%%%%%%%%
%%%%%%%%%%%%%%%%%%%%%%%%%%%%%%%%%%%%%
\begin{figure}[t]
\begin{center}
\includegraphics[width=9.0cm]{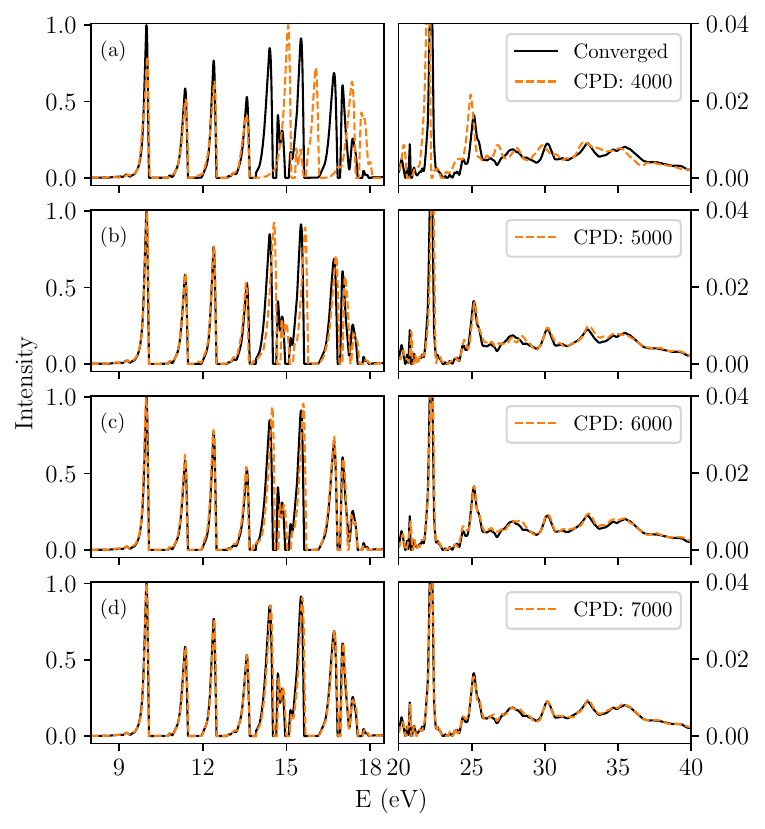}
\caption{Convergence of electronic ionization spectra of glycine with different
         CPD rank.
         The spectra in red are calculated with CPD-SQR
         Hamiltonian with 4000, 5000, 6000, and 7000 CPD ranks and shown in
         (a), (b), (c), and (d), respectively. Each plot's left and right panels 
         show the main peaks (8-18 eV region) and satellite peaks (20-40 eV regions), respectively.
         The numerically converged (regarding Hamiltonian terms) spectra are shown in black and calculated with
         10000 CPD rank. Further increases in the CPD rank do not change the spectrum.
         }
\label{fig:glycine_ip}
\end{center}
\end{figure}
%%%%%%%%%%%%%%%%%%%%%%%%%%%%%%%%%%%%%
Glycine is one of the smallest amino acids and it has been considered as an example to study
charge migration in biological systems using attosecond laser systems.
Its ionization spectrum has been studied in the past with the ADC(3) method. Therefore, we
consider it and calculate its ionization spectrum and real-time charge migration dynamics
using the MCTDH-SQR approach.
The cc-pVDZ basis is used for H, C, O, and N to generate the MOs, and we consider
20 electrons in
46 (20 highest occupied and 26 lowest unoccupied) spin orbitals in
our calculations.
The scheme of FS-DOFs and active spaces is shown in Table~\ref{tab:glycine_fsdof}.
The initial wavefunction for the propagation is generated by applying an ionization
operator that covers all occupied orbitals to the ground electronic state of
neutral glycine.
%The ionization operator reads
%%%%%%%%%%%%%%%%%%%%%%%%%%%%%%%%%%%%%
%\begin{align}
% \label{eq:iOpGlycine}
% \hat{A} = \sum_{i=1}^{10}\sum_{\sigma=\uparrow, \downarrow}
%           \hat{a}_{i,\sigma} .
%\end{align}
%%%%%%%%%%%%%%%%%%%%%%%%%%%%%%%%%%%%%
%This initial wavefunction is a spin doublet and overlaps with the singly
%ionized states of glycine.
The total number of terms in the original SQR
Hamiltonian is now $1.072\cdot 10^{6}$, which can be summed down to $1.11\cdot 10^{5}$ terms.
Fig.~\ref{fig:glycine_ip} compares the electronic ionization spectrum of
glycine using the CPD-SQR Hamiltonian with increasing ranks.
The spectrum in black corresponds to rank $10^{4}$ and is numerically converged.
Increasing the rank does not change the spectrum.
About 5000 to 6000 terms already provide a very accurate description and 4000 terms is
quite accurate in the challenging inner-valence region above 20~eV.

%%%%%%%%%%%%%%%%%%%%%%%%%%%%%%%%%%%%%
\begin{figure}[t]
\begin{center}
\includegraphics[width=9.0cm]{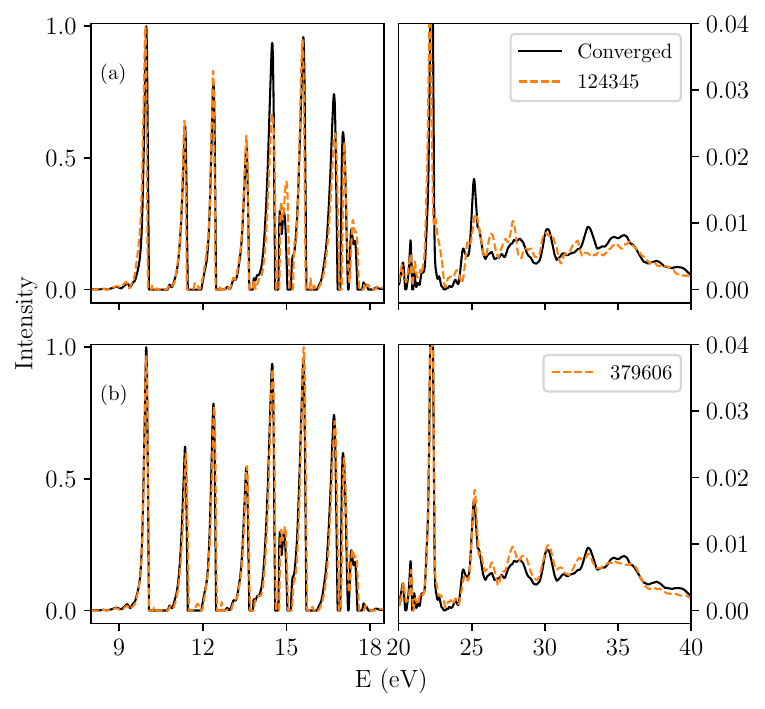}
\caption{Convergence of electronic ionization spectra of glycine with different
         propagated ML-MCTDH coefficients.
         The spectrum in black is calculated with the converged MCTDH wavefunction where
         the lowest natural population of each node of the MCTDH wavefunction is less than
         $5\cdot10^{-4}$.
         The spectra in red are calculated with 124345 and 379606 propagated
         ML-MCTDH coefficients and shown in 
         (a) and (b), respectively.
         }
\label{fig:glycine_wfn}
\end{center}
\end{figure}
%%%%%%%%%%%%%%%%%%%%%%%%%%%%%%%%%%%%%
In this example, the adopted FS-DOF scheme, as detailed in Table~\ref{tab:glycine_fsdof} generates
$\sim 8.7\cdot10^{9}$ primitive configurations. In comparison, the Hilbert space of $10$ alpha
and $10$ beta electrons within this Fock space comprises of
$\sim 6.2\cdot10^{8}$ configurations.
Furthermore, the numerically converged spectrum of ML-MCTDH is obtained by
approximately $5 \times 10^5$ propagated coefficients, highlighted in black in
Fig.~\ref{fig:glycine_wfn}. 
%To show the effect of the convergence of the ML-MCTDH wavefunction on the ionization spectrum, two more calculations are done with fewer propagated ML-MCTDH coefficients and compared with the converged spectrum in Fig.~\ref{fig:glycine_wfn}.
%It's clear from the figure that even with a significantly (more than 3 orders of magnitude) lower number of ML-MCTDH coefficients ($1.2\cdot 10^{5}$) compared to the number of CI coefficients ($\sim 6.2\cdot10^{8}$) one obtains a qualitatively good description of ionization spectrum.
Notably, the figure demonstrates that despite a substantial reduction
(by more than three orders of magnitude) in the number of ML-MCTDH coefficients
(approximately $1.2 \times 10^5$) compared to the number of CI coefficients
($\sim 6.2 \times 10^8$), a qualitatively accurate depiction of the ionization
spectrum is achieved.

%%%%%%%%%%%%%%%%%%%%%%%%%%%%%%%%%%%%%
\begin{figure}[t]
\begin{center}
\includegraphics[width=8.5cm]{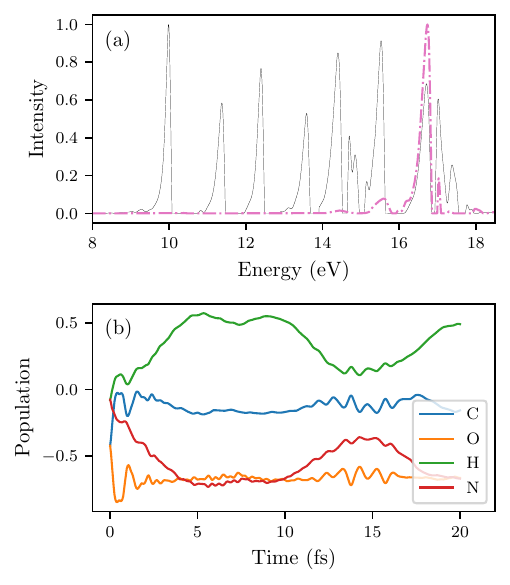}
\caption{Charge migration dynamics after removal of an electron from HOMO-8 molecular orbital.
         (a) shows the corresponding signature in the ionization spectra.
         (b) shows the atomic population dynamics with respect to the
         initial population of the constituting atoms of glycine.
         }
\label{fig:glycine_hole_2}
\end{center}
\end{figure}
%%%%%%%%%%%%%%%%%%%%%%%%%%%%%%%%%%%%%
In the MCTDH-SQR approach, one can naturally extract various time-dependent properties of the system.
For example, the ultrafast hole dynamics after the removal of electrons from the HOMO-8 spatial
orbitals is shown in Fig.~\ref{fig:glycine_hole_2}. The signature spectrum of the removal of an electron from
the HOMO-8 spatial orbital is shown in Fig.~\ref{fig:glycine_hole_2}(a).
Fig.~\ref{fig:glycine_hole_2}(b) shows the time-dependent electronic populations of C, O, H, and N atom
where the corresponding electronic population of the neutral ground state is subtracted. 
Therefore, the positive and negative values indicate the presence of extra electrons and holes with respect
to the neutral ground state, respectively.
These atomic populations are obtained from the MO populations directly available from the propagated
ML-MCTDH wavefunctions and their corresponding atomic orbital coefficients.
The total electronic population at each point in time is -1,
which indicates the presence of one hole in the system.
The population at $t=0$ reveals the atomic character of the MO
from which the electron is removed. In this case, the HOMO-8 molecular orbital
corresponds roughly to 42\%, 42\%, 8\%, and 8\% of C, O, H, and N characters, respectively.
The initial rapid jump in population in a subfemtosecond time-scale
is due to
the redistribution of the hole in the presence of an instantaneous
electron-electron repulsion~\cite{Ced99:205, Bre05:033901}.
Further, the removal of an electron from the HOMO-8 th orbital breaks down the orbital
picture of ionization, which is clear from the multiple peaks in the ionization spectrum
~\cite{ced86:115}.
After the rapid initial rearrangement, the atomic population of
carbon and oxygen remain more or less constant, and the
hole density hops between
nitrogen and hydrogen with a charge migration period of $\sim$ 15~fs.
After the first cycle, the H atoms have gained some net electron density
whereas the N atom has slightly increased its net hole density.

%%%%%%%%%%%%%%%%%%%%%%%%%%%%%%%%%%%%%
\begin{figure}[t]
\begin{center}
\includegraphics[width=8.5cm]{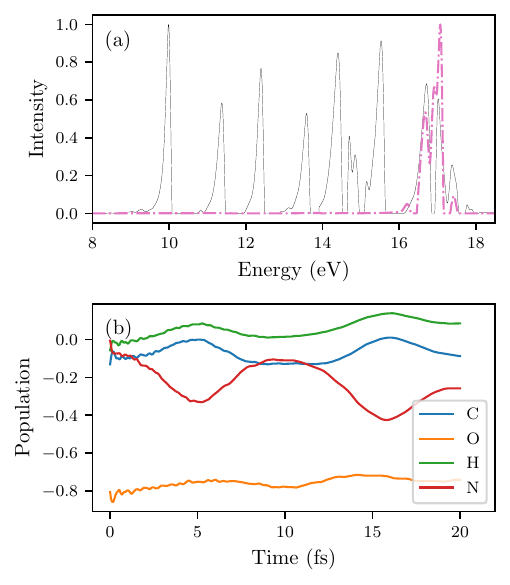}
\caption{Charge migration dynamics after removal of an electron from HOMO-9 molecular orbital.
         (a) shows the corresponding signature in the ionization spectra.
         (b) shows the atomic population dynamics with respect to the
         initial population of the constituting atoms of glycine.
         }
\label{fig:glycine_hole_1}
\end{center}
\end{figure}
%%%%%%%%%%%%%%%%%%%%%%%%%%%%%%%%%%%%%
On the other hand, a different hole dynamics is observed if one removes an electron from HOMO-9
molecular orbital (shown in Fig.~\ref{fig:glycine_hole_1}) and
corresponding to an ionization potential of about 17~eV.
The HOMO-9 MO is spread to about
13\%, 80\%, 6\%, and 1\% onto the C, O, H, and N atoms,
respectively.
The electronic population at the oxygen atom (where most of the hole lies)
remains more or less constant.
The rest of the hole density hops between the N atom
and its adjacent C and H atoms with a period of about 10~fs.
After each cycle, the N atom loses some electronic population and C and H atoms slowly gain electronic population.
%%%%%%%%%%%%%%%%%%%%%%%%%%%%%%%%%%%%%

These dynamics are susceptible to being probed by ultrafast pump-probe
schemes in the attosecond regime. Other approaches such as the ADC(3) method
for ionization can provide equivalent information in the
frequency domain~\cite{Sch83:1237, Sch98:4734}. An interesting breakthrough for \emph{ab initio}
MCTDH-SQR could be achieved by devising a scheme to consider nuclear displacements
at early-times after ionization.
This has been demonstrated for
one single nuclear coordinate and a
simpler SQR
Hamiltonian in Ref.~\citenum{Sas20:154110}, and its multidimensional extension
shall be the subject
of future investigations.

%%%%%%%%%%%%%%%%%%%%%%%%%%%%%%%%%%%%%%%%%%%%%%%%%%%%%%%%%%%%%%%%%%%%%%%%
%%%%%%%%%%%%%%%%%%%%%%%%%%%%%%%%%%%%%%%%%%%%%%%%%%%%%%%%%%%%%%%%%%%%%%%%
\section{Scaling of the ML-MCTDH-SQR method}  \label{sec:scaling}
%%%%%%%%%%%%%%%%%%%%%%%%%%%%%%%%%%%%%%%%%%%%%%%%%%%%%%%%%%%%%%%%%%%%%%%%
%%%%%%%%%%%%%%%%%%%%%%%%%%%%%%%%%%%%%%%%%%%%%%%%%%%%%%%%%%%%%%%%%%%%%%%%
%%%%%%%%%%%%%%%%%%%%%%%%%%%%%%%%%%%%%
\begin{figure}[t]
\begin{center}
\includegraphics[width=8.0cm]{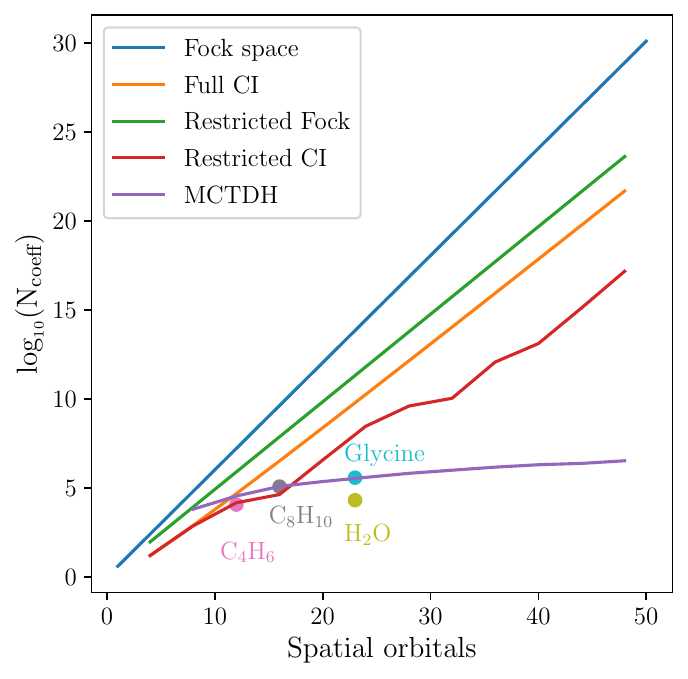}
\caption{Scaling of various coefficients related to the ML-MCTDH-SQR
         method with the number of spatial orbitals.
         The solid lines represent the projected number of coefficients
         taking into account
         the underlying assumptions on the wavefunction (see text).
         The circles represent the total number of
         propagated ML-MCTDH coefficients, color-coded to denote
         different numerical examples presented in the text.
         See supplementary material for the calculation of
         C$_4$H$_6$ system.
         }
\label{fig:mctdhsqr_coeff}
\end{center}
\end{figure}
%%%%%%%%%%%%%%%%%%%%%%%%%%%%%%%%%%%%%

Up to now, we focused on the compactification
of the CPD-SQR operator. Equally important
is the question of how much
compactification is achieved for the
wavefunction with respect to the primitive
configurational space.
%The results are shown in Fig.~\ref{fig:mctdhsqr_coeff}.
%
The MCTDH-SQR primitive space is
spanned, in principle, by the
$2^{2N}$ configurations in the
total Fock space, where
N is the number of spatial orbitals (blue line in Fig.~\ref{fig:mctdhsqr_coeff}).
%
%However, we have already discussed how, in practice, one defines an active space within each Fock sub-space determined by its maximal and minimal occupation, which accounts in practice to a static pruning of the primitive degrees of freedom.
However, as previously discussed, practical implementation involves defining an
active space within each Fock sub-space based on its maximum and minimum
occupation.
This effectively constitutes a static pruning of the primitive degrees of freedom.
The number of configurations in the
selected Fock space is shown in green in
Fig.~\ref{fig:mctdhsqr_coeff}.
In practice, we often consider sub-Fock spaces formed by
four spatial (eight spin) orbitals with an
occupation that varies
by $\pm3$ electrons with respect to their reference
occupation.
The so-selected Fock space corresponds to the actual primitive space (or
primitive grid) of the MCTDH calculation.
If we consider now that the total
occupation
of the molecular orbital space is equal to
$N/4$, i.e., 75\% of the orbitals are
virtual in the reference, we can determine
the number of configuration interaction
(CI) basis states within the corresponding number of electrons.
The CI states arising from the full Fock space make up the full CI space shown in orange.
The CI states arising from the selected
Fock space build up the selected CI space,
shown in red. The latter is the important number
to be compared with the total number of
propagated MCTDH-SQR coefficients.
In other words, the primitive space that
the MCTDH-SQR tensor-tree spans corresponds to the green trace, but the
physical CI space corresponding to a specific number of electrons
corresponds to the red trace.
Only if the total number of
propagated MCTDH-SQR coefficients is
substantially lower than the number of selected
CI states it is meaningful to undertake
an MCTDH-SQR calculation.
If they are similar, and it is affordable,
it is certainly more efficient
to construct the selected CI Hamiltonian
and work with it directly.

We have to assume finally how to structure the layering of the ML-MCTDH-SQR
wavefunction and the number of SPFs in each layer:
we consider balanced trees (as symmetric as possible), where each parent node
connects to a maximum of two child nodes (three for the top layer).
The number of SPFs in the bottom-layer nodes is set as $1/3$ of the primitive basis size,
and the SPF count in a parent node is the sum of the SPFs in its child nodes.
Consequently, the number of SPFs increases linearly with the layer of the MCTDH tree.
This number as a function of $N$ is shown in purple in
Fig.~\ref{fig:mctdhsqr_coeff}.
This estimate indicates that for about 20 spatial orbitals and above,
the number of ML-MCTDH-SQR coefficients is several orders of magnitude smaller
than the number of Hilbert space configurations in the selected CI space of the
corresponding number of electrons. This gain increases quickly with the number of spatial
orbitals. Crucially, this estimate suggests that an \emph{ab initio}
ML-MCTDH-SQR approach succeeds at compactifying the full-CI state of the system
while starting from a much larger Fock space. We note here that, in the ML-MCTDH-SQR
approach, no effort whatsoever is made to enforce the wavefunction to the Hilbert space
of the problem. Instead, the equations of motion conserve the total number of particles
automatically, and this observable becomes \emph{de facto} a convergence parameter.

%%%%%%%%%%%%%%%%%%%%%%%%%%%%%%%%%%%%%%%%%%%%%%%%%%%%%%%%%%%%%%%%%%%%%%%%
%%%%%%%%%%%%%%%%%%%%%%%%%%%%%%%%%%%%%%%%%%%%%%%%%%%%%%%%%%%%%%%%%%%%%%%%
\section{Summary and Conclusion}  \label{sec:conclusions}
%%%%%%%%%%%%%%%%%%%%%%%%%%%%%%%%%%%%%%%%%%%%%%%%%%%%%%%%%%%%%%%%%%%%%%%%
%%%%%%%%%%%%%%%%%%%%%%%%%%%%%%%%%%%%%%%%%%%%%%%%%%%%%%%%%%%%%%%%%%%%%%%%
In this work, we introduce a sum-of-products (SOP) representation for the
electronic Hamiltonian within the framework of second quantization (SQR). 
The quartic scaling of the electronic SQR Hamiltonian with respect to the
number of spin orbitals presents a significant challenge when addressing
the time-dependent phenomena in large molecular systems.
To tackle this issue, we propose a method based on the canonical polyadic
decomposition (CPD) to obtain a compact sum-of-product form of the electronic
Hamiltonian. In contrast to the Tucker decomposition, the basis functions
of a CPD need not be orthogonal to each other, which allows it to create
a more compact representation.
However, unlike multidimensional potential energy surfaces,
the Hamiltonian tensor is highly discontinuous along its dimensions,
making it unsuitable for employing a Monte Carlo integration scheme
to obtain the CPD form of the Hamiltonian. Here we introduce a practical
scheme (SOP to SOP) to obtain the CPD representation of the Hamiltonian.
We benchmark our method on H$_2$O,
a trans-polyene, and the ultrafast inner-valence dynamics of ionized
glycine, where we demonstrate the compactness of the CPD-SQR Hamiltonian in comparison to
the exact Hamiltonian. In particular, CPD-SQR achieves numerical convergence
within the spectral resolution of the calculated spectra
with a CPD rank of more than
two orders of magnitude smaller than the original
SQR Hamiltonian.
Although the various examples indicate that the accuracy of the state
energies improves systematically with the CPD rank,
the total accuracy remains limited by the wavefunction propagation time
and the corresponding duration of the autocorrelation function.
The spectral width in the chosen examples is about 0.1~eV FWHM.
We emphasize that our aim with the ML-MCTDH-SQR method and the CPD representation of
the operator is to have a variational approach that combines efficiency and robustness
to simulate broad spectral
ranges for short times, potentially covering regions with very high densities of electronic states (cf. glycine example),
whereas high precision and accuracy of single states are of less importance. 
%
%We investigate the charge migration dynamics for some
%inner-valence ionized states
%of glycine and compare the corresponding ionization spectrum, obtained in the
%time domain, and the
%real-time dynamics of the electrons.

A program (named \texttt{compactoper}) implementing the algorithm discussed in
this work has been made available in the Heidelberg MCTDH
package~\cite{mctdh:MLpackage}.
%%%%%%%%%%%%%%%%%%%%%%%%%%%%%%%%%%%%%%%%%%%%%%%%%%%%%%%%%%%%%%%%%%%%%%%%
\section{Supplementary Material}  \label{sec:supple}
%%%%%%%%%%%%%%%%%%%%%%%%%%%%%%%%%%%%%%%%%%%%%%%%%%%%%%%%%%%%%%%%%%%%%%
See the supplementary material for the excitation spectrum calculation of
trans-C$_4$H$_6$ system.

%%%%%%%%%%%%%%%%%%%%%%%%%%%%%%%%%%%%%%%%%%%%%%%%%%%%%%%%%%%%%%%%%%%%%%
\section{Data Availability}  \label{sec:data}
%%%%%%%%%%%%%%%%%%%%%%%%%%%%%%%%%%%%%%%%%%%%%%%%%%%%%%%%%%%%%%%%%%%%%%
The data that support the findings of this study are available
within the article and its supplementary material.
A program (named \texttt{compactoper}) implementing the algorithm discussed in
this work has been made available in the Heidelberg MCTDH
package~\cite{mctdh:MLpackage}.
%%%%%%%%%%%%%%%%%%%%%%%%%%%%%%%%%%%%%%%%%%%%%%%%%%%%%%%%%%%%%%%%%%%%%%

%%%%%%%%%%%%%%%%%%%%%%%%%%%%%%%%%%%%%%%%%%%%%%%%%%%%%%%%%%%%%%%%%%%%%%%%
\section{Acknowledgments}
%%%%%%%%%%%%%%%%%%%%%%%%%%%%%%%%%%%%%%%%%%%%%%%%%%%%%%%%%%%%%%%%%%%%%%%%
We thank Prof. H.-D. Meyer for the insightful discussions and
advice with the MCTDH calculations.
We thank Prof. Henrik R. Larsson
for discussions on the
diagrammatic description of operators.
The authors thank JUSTUS 2 in Ulm under grant number bw18K011 for computing time.
The authors declare no conflicts of interest.

%\appendix
%\section{}

%\bibliography{refs,cites}
%\bibliographystyle{aip}
%\bibliography{cites}
%\begin{thebibliography}{99}
%\end{thebibliography}

%aipnum4-2.bst 2019-01-14 (MD) hand-edited version of apsrev4-1.bst

\end{document}